\documentclass[11pt]{article}

\usepackage{endfloat,mathrsfs}
\usepackage{amsmath,amsfonts}
\usepackage{multirow}
\usepackage{color}
\usepackage{dashrule}
\usepackage{listings}
\usepackage{bm}

\usepackage[default]{jasa_harvard}    
\usepackage{JASA_manu}

\renewcommand{\harvardand}{and}

\newcommand{\ba}{\begin{array}}
\newcommand{\ea}{\end{array}}

\newtheorem{theorem}{Theorem}

\newtheorem{lemma}[theorem]{Lemma}
\newenvironment{proof}{\paragraph{Proof:}}{\hfill$\square$}

\begin{document}

\title{A marginalizable frailty model for correlated right-censored data}

\author{Rui Zhang  \\ and \\Kwun Chuen Gary Chan    \\  \\
Department of Biostatistics \\ University of Washington, Seattle, WA 98195  \\
emails: \texttt{zhangrui@u.washington.edu}\\      \texttt{kcgchan@u.washington.edu}}

\maketitle

\newpage
\begin{center}
\textbf{Abstract}
\end{center}
We introduce a flexible individual frailty model for clustered right-censored data, in which covariate effects can be marginally interpreted as log failure odds ratios. Flexible correlation structures can be imposed by introducing multivariate exponential distributed frailties, constructed from a set of multivariate Gaussian random variables. Finite and infinite dimensional parameters are consistently estimated by maximizing a composite contributing marginal likelihood and a consistent estimate for their asymptotic covariance is proposed. Parameter estimation is implemented through a hybrid expectation-maximum algorithm. Simulations and an analysis of the Rats study were carried out to demonstrate our method. 

\vspace*{.3in}


\newpage

\section{Introduction}
\citeasnoun{ref52} proposed the proportional hazards model for analyzing right-censored survival data; this model, along with its generalizations, are the most popular models in this field. Hazard rates from different sub-populations are assumed to be proportional over time. As for model inference, Cox (1972, 1975) proposed the elegant partial likelihood approach; corresponding estimates were shown to achieve the optimal semi-parametric estimation efficiency and a straightforward estimator of its asymptotic covariance was also proposed.

Alternative models are also discussed, partly due to the limitations imposed by the time-invariant proportional hazards assumption. \citeasnoun{ref95} generalized the proportional odds model from categorical data into survival data. Given a covariate $Z$, the log survival odds at time $t$ is
\[
-\mathrm{logit}\left\{S(t\mid Z)\right\}=G(t)+Z^T\beta,\quad\text{where } \mathrm{logit}(x):=\mathrm{log}\left(\frac{x}{1-x}\right)\ .
\]
$G(t)$ is the log failure odds function at zero covariate value, i.e. the baseline. Hazard rates of sub-populations converge as time goes to infinity. \citeasnoun{ref95} suggested using this model for some effective cure with which the mortality rate in a diseased group is speculated to converge to its counterpart from the control group. \citeasnoun{ref90} proposed the non-parametric maximum likelihood estimation (NPMLE) inference method for independent data: the baseline log failure odds function is estimated non-parametrically as a non-decreasing and right-continuous step function, and the estimates are consistent and achieve the optimal semi-parametric estimation efficiency.

For familial-clustered datasets, observations are typically correlated. To account for dependence, \citeasnoun{ref103} introduced the notion of shared relative-risk between subjects from the same cluster; \citeasnoun{ref104} incorporated covariates. \citeasnoun{ref103} pointed out his model is equivalent to combining some cluster-common unobservable effects multiplicatively into the hazard rates in the Cox proportional hazards model, named frailties by \citeasnoun{ref33}. Further assuming observations to be independent conditional on frailties, this class of models are named frailty models. A frailty model is equivalent to a mixed effects model with random intercepts and a complementary-log-log conditional link. Shared frailty models can incorporate diverse correlation structures in twin-clustered studies: \citeasnoun{ref93} pointed out many bivariate survival distributions uniquely correspond to some shared frailty models. However, when there are more than two observations per cluster, the exchangeable correlation structure in shared frailty models is not sufficient. For example, in family-clustered studies, people usually use frailties to model latent genetic factors; however, a mono-zygotic twin pair definitely shares more genetic materials than a parent-offspring pair and thus are more correlated. \citeasnoun{ref37} proposed an individual frailty model with log-normal distributed frailties to allow flexible modeling of correlation.

In generally we cannot estimate \textit{marginal} covariate effects from frailty models. Yet marginal effects are directly interpretable and preferred to answer public health questions, according to \citeasnoun{book4} and \citeasnoun{ref7}. \citeasnoun{ref109}, \citeasnoun{ref48} considered multivariate failure time data and proposed a marginal Cox proportional hazards model. They developed an estimating equation with the same flavor as in \citeasnoun{ref106}, and used an independent working correlation structure. It is shown their estimates of marginal effects are consistent under correct marginal models. Other researchers used the underlying correlation structure to improve estimation efficiency, such as in \citeasnoun{ref74} and \citeasnoun{ref76}. The primary focus of these models is evaluating marginal covariate effects, while the underlying correlation level is usually of secondary interest or simply ignored.

Marginalizable frailty models can estimate population level covariate effects, and simultaneously model the dependence structure more straightforwardly by frailties as well as make predictions of latent cluster effects, i.e. frailties. Besides, since frailty models specify the full distributions, parameters of interest always exist, unlike the case in \citeasnoun{ref109}. \citeasnoun{ref78} discussed the use of shared frailties under a positive stable distribution. Marginally his model remains to be a Cox proportional hazards model with scaled covariate effects. Extensions were studied by \citeasnoun{ref107}, \citeasnoun{ref108} and \citeasnoun{ref102}. However, a stable distribution usually models the asymptotic behaviour of the partial sum of random variables without the first or second moments,
while such distributions are not widely used in biomedical applications. Besides, only shared frailty models are discussed, giving relatively rigid correlation structures.

A popular frailty model for correlated survival data assume Gamma distributed frailties, mainly for inference simplicity due to conjugacy between Gamma distribution and the complementary-log-log link. \citeasnoun{ref39} proposed the expectation-maximization (EM) algorithm for inference, in which the complete likelihood is constructed by including unknown Gamma distributed frailties. In the E-step, frailties are Gamma distributed with different parameters, greatly simplifying calculations. The M-step is equivalent to solving a partial score equation, offset by imputed frailties from the E-step. Respective asymptotic theories, such as consistency, asymptotic normality of the estimates and the corresponding likelihood ratio test, have been well studied by \citeasnoun{ref36} and \citeasnoun{ref96} for the cases with and without covariates. \citeasnoun{ref112} also discussed the behavior of NPMLE under more general conditions where the fitted model may be misspecified. On the other hand, other frailty distributions generally require heavy computation. The frailty model with individual log-normal frailties by \citeasnoun{ref37} adopts to a Monte Carlo method since there is no closed-form likelihood to work with. Penalized log-likelihood is an alternative: frailties are estimated as fixed effects, and to account for the high-dimensional problem, a penalty function composed by frailties is added into the penalized log-likelihood for stabilization. See \citeasnoun{ref92} for related discussions. Parameters in these models, however, do not admit marginal interpretations.

We propose a marginalizable individual frailty model for clustered survival datasets in Section 2, assuming fraities following a multivariate exponential distribution. With individual frailties, the model can allow a flexible correlation structure. Also, the model parameters have a population-level interpretation. In Section 3, we generalize NPMLE for independent data into composite likelihood as the non-parametric maximum composite likelihood estimate (NPMCLE), avoiding high-dimensional integrals. We provide a hybrid and tractable inference algorithm, composed of an EM algorithm adopted to composite likelihood for estimating marginal covariate effects and the baseline failure odds function, and a direct composite likelihood maximization for correlation parameters. Section 4 presents theorems regarding consistency and asymptotic normality of our estimates, together with a consistent variance estimator of these estimates. Subsection 5$\cdot$1 contains simulation results to evaluate the finite sample performance of our estimator. In Subsection 5$\cdot$2 we analyzed a Rats dataset from \citeasnoun{ref98} using the proposed model, followed by discussions of scientific explanations and justifications for our frailty model. Section 6 covers concluding remarks and discussions. Technical details are discussed in the appendix.

\section{Model}
\subsection{Notations}
Our model is designed for clustered right-censored survival datasets and the goal is to estimate the association between the response variable, i.e., the survival time $T$, and a covariate vector $Z$. Survival times from the same cluster are correlated due to unobservable genetic, environmental or other factors, which are called frailties and denoted by $W$. Like most frailty models, we assume $W$ fully explain the correlation not captured by the covariates.

Consider a clustered dataset composed by $m$ independent clusters, and from each cluster $i$, there are $n_i$ subjects sampled. We denote failure times by $(T_{i1},\ldots,T_{in_i})$ and censoring times by $(C_{i1},\ldots,C_{in_i})$. We observe right-truncated times $Y_{ij}=\mathrm{min}\left\{T_{ij}, C_{ij}\right\}$ and record failure indicator variables
\[
\Delta_{ij}=\left\{\ba{cc}1&\quad T_{ij}\leq C_{ij}\\  
0&\quad T_{ij} > C_{ij}\ea\right.\ ,
\]
along with covariate vectors $(Z_{i1},\ldots,Z_{in_i})$.
Denote frailties from the $i$th cluster as $(W_{i1},\ldots,W_{in_i})$. To summarize, each observation is denoted by $X_{ij}=(Y_{ij},\Delta_{ij},Z_{ij})$; observations from the $i^{th}$ cluster are denoted by 
\[
O_i=\{(Y_{ij},\Delta_{ij},Z_{ij}); j=1,\ldots,n_i\}=\{X_{ij}; j=1,\ldots,n_i\}\ .
\]

\subsection{Model}

For subject $j$ from cluster $i$, given its frailty $W_{ij}$ and the observed covariate vector $Z_{ij}$, we assume its conditional hazard rate takes the form of
\[
\lambda_{ij}(t\mid W_{ij}=w_{ij},Z_{ij}=z_{ij})=\lambda_0(t)w_{ij}\mathrm{exp}\left(z_{ij}^T\beta\right), \quad t>0,
\]
where $\lambda_0(\cdot)$ is the conditional baseline hazard rate function at zero-value covariate and $\beta$ stands for the conditional log hazard rates ratio. We assume $W_{ij}$'s are marginally standard exponential distributed, but are correlated as discussed in Subsection 2$\cdot$3.

To marginalize the conditional model, note that
\[
S(t\mid Z_{ij}=z_{ij}): =\int_0^\infty S_{ij}(t\mid z_{ij},w_{ij})f(w_{ij})dw_{ij}=
\left\{1+\Lambda_0(t)\mathrm{exp}\left(z_{ij}^T\beta\right)\right\}^{-1}\ ,
\]
i.e. odds of failure is
\begin{equation}\label{marginal.survival}
\frac{1-S(t\mid z_{ij})}{S(t\mid z_{ij})}=\Lambda_0(t)\mathrm{exp}\left(z_{ij}^T\beta\right)\ .
\end{equation}
Therefore, $\beta$ can be interpreted as the marginal proportional failure log odds ratio. Although the exponential distributed frailty is a special case of the Gamma family, we consider an individual frailty $W_{ij}$ instead of a shared frailty, and this allows flexible modeling of correlation structure. Further discussion of the exponential frailty assumptions are given in Section 6.

\subsection{Multivariate exponential distributed frailties}
We assume individual frailties from the same cluster follow a multivariate standard exponential distribution, which is a special case of the multivariate Gamma distribution. Generation of a multivariate Gamma random vector from a set of independently and identically distributed (i.i.d.) normal vectors was discussed by \citeasnoun{ref29}, and \citeasnoun{ref18}. Let $V_1, V_2\in\mathbb{R}^p$ be two i.i.d. mean zero p-variate normal random vectors, each can be written as $V_j=(V_{j1},\ldots,V_{jp})$, $j=1,2$, with a $p\times p$ covariance matrix $C$ having unit diagonal elements. Set $W_d=(V_{1d}^2+V_{2d}^2)/2$, $d=1,\ldots,p$; $2W_d$ is marginally $\chi^2(2)$ distributed, or equivalently $W_d \sim \mathrm{Exp}(1)$. Moreover, the correlation matrix of $(W_1,\ldots,W_p)$, $R(\rho)$ where $\rho$ can be a vector-valued parameter, is an element-wise square of $C$, as discussed in \citeasnoun{ref18}. Since this particular class of multivariate exponential distribution is closely related with the multivariate normal distribution, it can be used to model highly flexible positive correlation structures.

\subsection{Frailty Dispersion and Covariate Dependence}

Consider the case where frailties marginally follow exponential distribution with parameter $\eta$. Marginalizable property of our model remains yet $\eta$ merges with the intercept $\lambda_0$ and thus is non-identifiable. To see this, consider rescaled frailties $\tilde{W}_{ij}:=\eta W_{ij}$,
\[
\lambda_{ij}(t\mid Z_{ij}=z_{ij},W_{ij}=w_{ij})=\lambda_{0}(t)\mathrm{exp}\left(-w_{ij}e^{-z_{ij}^T\beta}\right)
=\tilde{\lambda}_{0}(t)\mathrm{exp}\left(-\tilde{w}_{ij}e^{-z_{ij}^T\beta}\right), \quad\text{where }\tilde{\lambda}_0(\cdot)=\left\{\lambda_0(\cdot)\right\}^{1/\eta}\ .
\]
Therefore, we can standardize the frailty dispersion to be one. We also consider the case where frailty dispersions depend on covariates. A similar situation is discussed by \citeasnoun{ref102} in a study comparing the ratio of observed to expected deaths, known as the standardized mortality ratio, between U.S. kidney transplant centers and the national average. Suppose given a covariate $Z_{ij}$, $W_{ij}$ is marginally exponential distributed with mean $\mathrm{exp}(Z_{ij}^T\eta)$ and a correlation matrix $R(\rho)$. Consider re-scaled frailties $\tilde{W}_{ij}:=e^{-Z_{ij}^T\eta}W_{ij}$,
\[
\lambda_{ij}(t\mid Z_{ij}=z_{ij},W_{ij}=w_{ij})=\lambda_{0}(t)\mathrm{exp}\left(-w_{ij}e^{-z_{ij}^T\beta}\right)
=\lambda_{0}(t)\mathrm{exp}\left(-\tilde{w}_{ij}e^{-z_{ij}^T\tilde{\beta}}\right), \quad\text{where }\tilde{\beta}=\beta-\eta\ .
\]
The marginal odds of failure in \eqref{marginal.survival} becomes
\[
\frac{1-S(t\mid z_{ij})}{S(t\mid z_{ij})}=\Lambda_0(t)\mathrm{exp}\left(z_{ij}^T\tilde{\beta}\right)\ .
\]
Still, marginalizable property of our model holds and the marginal parameter $\tilde{\beta}$ can be estimated as if the frailties were covariate independent.

\subsection{Scientific justification of frailties}
Gamma distributed frailties are chosen in most frailty model literature due to mathematical convenience, but here we give out a physical justification for the exponential frailties. The logarithm of frailty is Gumbel distributed, which can model the limiting distribution of the maximum from a set of normal or exponential type random variables. Thus exponential frailties are reasonable when we believe there are many latent individual effects and the maximum dominates the others in affecting the outcome.

\section{Model inference: A hybrid generalized EM algorithm}
\subsection{Composite contributing marginal log-likelihood}
Denote the parameters under interest as $\theta:=(\beta,\rho,\Lambda)$, since $\Lambda$ can be estimated at the same rate as the other finite-dimensional parameters $(\beta,\rho)$. The true parameter is denoted by $\theta_0:=(\beta_0,\rho_0,\Lambda_0)$. In this section we propose an inference method for $\theta$, extending the work by \citeasnoun{ref90} for independent data. 

First, we need to find some likelihood to work with. The following conditions are for the identifiability of our model and the construction of the contributing marginal log-likelihood:\\
C1. Given frailties $W_{i}$, the distribution of covariate vectors $Z_i$ is independent of frailties and is non-informative, i.e. it does not contain $\theta$.\\
C2. (Coarsening at random assumption) Conditioning on $(Z_{i},T_{i},W_i)$, the hazard rate function of censoring time $C_{ij}$ is only a function of covariates $Z_i$ and is non-informative.\\
C3. The distribution of cluster observation counts is independent of censoring and death times, and the number of observations in each cluster is uniformly bounded by $n_0$.\\
C4. The true conditional baseline cumulative hazard function $\Lambda_0(t)$ is a strictly increasing function on $[0,\tau]$ and is continuously differentiable. In addition, $\Lambda_0(0)=0$ and $\Lambda_0'(0)>0$.\\
C5. Parameter spaces of $\beta$ and $\rho$, denoted by $\mathcal{B}$ and $\mathcal{R}$, belong to some known convex and compact subsets of $\mathbb{R}^{p_1}$ and $\mathbb{R}^{p_2}$, respectively:
\begin{eqnarray*}
\mathcal{B}&:=&\left\{\beta\in \mathbb{R}^{p_1}:||\beta|| \leq B_0\text{ for some finite constant $B_0$}\right\}\ ,\\
\mathcal{R}&:=&\left\{\rho\in \mathbb{R}^{p_2}:||\rho|| \leq R_0\text{ for some finite constant $R_0$}\right\}\ ,
\end{eqnarray*}
where $||\cdot||$ denotes the Euclidean norm; the true value $(\beta_0,\rho_0)$ is not a boundary point of $\mathcal{B}\times\mathcal{R}$.\\
C6. $Z$ is not degenerate; i.e. $Z^T\beta=0$ a.s. implies $\beta=0$ $F_Z$-almost surely, where $F_Z$ is the joint distribution of covariate vectors. 

Conditions C1 and C3 guarantee that inference can be made while ignoring the covariate and cluster observation generation distributions. The assumption in C2 follows from \citeasnoun{ref39} and \citeasnoun{ref91} as a non-informative censoring condition. Condition C4 removes the cases of ties and together with C6, leading to model identifiability. Condition C5 is an usual technical condition for parameter spaces. 

It follows from C1-C3 that we can first remove factors involving covariate, censoring time and observation generation procedure distributions from the full likelihood and integrate over frailties, getting the marginal contributing log-likelihood for observation $O_i$: 
\begin{scriptsize}

\begin{eqnarray}
&&\mathrm{log}\int_{w_{in_i}}\cdots\int_{w_{i1}}\left\{\overset{n_i}{\underset{j=1}{\prod}}\left[ f_T(T_{ij}=y_{ij}\mid z_{ij},w_{ij})\right]^{\Delta_{ij}}\left[\mathrm{pr}(T_{ij}>y_{ij}\mid z_{ij},w_{ij})\right]^{1-\Delta_{ij}}
\right\}\times f_{W_i}(w_{i};\rho)dw_{i1}\cdots dw_{in_i}\nonumber\\=&&\overset{n_i}{\underset{j=1}{\sum}}\Delta_{ij}\left[\mathrm{log}\,\lambda_0(y_{ij})+Z_{ij}^T\beta\right]
+ \mathrm{log}\int_{w_{in_i}}\cdots\int_{w_{i1}}\left\{\overset{n_i}{\underset{j=1}{\prod}}w_{ij}^{\Delta_{ij}}
\mathrm{pr}(T_{ij}>y_{ij}\mid z_{ij},w_{ij})
\right\}\times f_{W_i}(w_{i};\rho)dw_{i1}\cdots dw_{in_i}\label{survival_joint_marginal}
\end{eqnarray}
\end{scriptsize}
where $f_T$ stands for the conditional density of failure time. This is equivalent to the contributing marginal likelihood, obtained by first integrating over frailties then removing irrelevant terms, as discussed by \citeasnoun{ref39} and \citeasnoun{ref83}.

Integration in \eqref{survival_joint_marginal} uses the Laplace transformation of frailties:
\[
\mathscr{L}_{W_i}(u_i):=\mathrm{E}_{W_i}\left[\mathrm{exp}\left( -\underset{j=1}{\overset{n_i}{\sum}}w_{ij} \Lambda(y_{ij}) e^{z_{ij}^T\beta} \right)\right]=
|I+C\text{diag}(  \Lambda(y_{i1})e^{z_{i1}^T\beta},\ldots, \Lambda(y_{in_i})e^{z_{in_i}^T\beta})|^{-1}\ ,
\]
where $u_i:=\left(\Lambda(y_{i1})e^{z_{i1}^T\beta},\ldots,\Lambda(y_{in_i})e^{z_{in_i}^T\beta}\right)$, and $C$ is the element-wise square root of $(W_{i1},\ldots,W_{in_i})$ correlation matrix $R(\rho)$. Integral in \eqref{survival_joint_marginal} is proportional to:
\[
\overset{n_i}{\underset{j=1}{\prod}}\left\{\lambda(y_{ij})\mathrm{exp}\left(z_{ij}^T\beta\right)
\right\}^{\delta_{ij}}\cdot\frac{\partial ^{d_i}\mathscr{L}_{W_i}(u_i)}{\overset{n_i}{\underset{j=1}{\prod}}\left(\partial u_{ij}\right)^{\delta_{ij}}},\quad \text{where }d_i:=\overset{n_i}{\underset{j=1}{\sum}}\delta_{ij}\ .
\]
We are taking partial derivatives of this Laplace transformation at values corresponding to failure events. Since $\mathscr{L}$ is in the form of a matrix determinant inverse, it is quite complicated to work with. Thus, we chose to work with a composite contributing marginal log-likelihood, which is a weighted summation of pairwise contributing marginal log-likelihoods. This is equivalent to working with a mis-specified model under which correlations among three or more observations are ignored. We will show the resulting estimator is still consistent in the next section, but the estimation efficiency is partially sacrificed since we are not working with the true joint likelihood.

Here we explicitly write out a pairwise contributing marginal log-likelihood. Denote
\begin{eqnarray*}
u(X_{ij};\beta,\Lambda)&=&\Lambda(Y_{ij})e^{Z_{ij}^T\beta}\ ,\\
v(X_{ij},X_{ik};\beta,\rho,\Lambda)&=&(1-\rho_{jk})u(X_{ij};\beta,\Lambda)u(X_{ik};\beta,\Lambda)+u(X_{ij};\beta,\Lambda)+u(X_{ik};\beta,\Lambda)+1\ ,\\
w(X_{ij},X_{ik};\beta,\rho,\Lambda)&=&\Delta_{ij}\Delta_{ik}(1-\rho_{jk})^2u(X_{ij};\beta,\Lambda)u(X_{ik};\beta,\Lambda)\\&&+\Delta_{ij}(1-\rho_{jk})u(X_{ik};\beta,\Lambda)+\Delta_{ik}(1-\rho_{jk})u(X_{ij};\beta,\Lambda)+1+\Delta_{ij}\Delta_{ik}\rho_{jk}\ ,
\end{eqnarray*}
where $\rho_{jk}$ is the correlation between $W_{ij}$ and $W_{ik}$. In the case of an exchangeable correlation structure, $\rho_{jk}$'s identically equal to a scalar parameter $\rho$. In general, $\rho_{jk}$ is a function of $\rho$, which can be a vector.

The pairwise contributing marginal log-likelihood of correlated observations $(X_{ij},X_{ik})$ is:
\begin{eqnarray}
l(X_{ij},X_{ik};\beta,\rho,\Lambda)&=&\mathrm{log}w(X_{ij},X_{ik};\beta,\rho,\Lambda)
+\Delta_{ij}\mathrm{log}\lambda(Y_{ij})+\Delta_{ik}\mathrm{log}\lambda(Y_{ik})
\nonumber\\&&+(\Delta_{ij}Z_{ij}+\Delta_{ik}Z_{ik})^T\beta
-(1+\Delta_{ij}+\Delta_{ik})\mathrm{log}v(X_{ij},X_{ik};\beta,\rho,\Lambda)\ .\label{composite}
\end{eqnarray}
The composite contributing marginal log-likelihood of $O_i$ is
\begin{scriptsize}
\begin{eqnarray}
&&clog(O_i;\beta,\rho,\Lambda)\nonumber\\=&&\frac{1}{n_i-1}{\underset{j<k}{\sum}}l(X_{ij},X_{ik};\beta,\rho,\Lambda)\nonumber\\=&&
\underset{j=1}{\overset{n_i}{\sum}}\Delta_{ij}\left\{\mathrm{log}\lambda(Y_{ij})+Z_{ij}^T\beta\right\}
+\frac{1}{n_i-1}{\underset{j<k}{\sum}}\left\{\mathrm{log}w(X_{ij},X_{ik};\beta,\rho,\Lambda)    -(1+\Delta_{ij}+\Delta_{ik})\mathrm{log}v(X_{ij},X_{ik};\beta,\rho,\Lambda)\right\}\ .\label{NPMCLE_motive1}
\end{eqnarray}
\end{scriptsize}
Switching to a composite log-likelihood, the high-dimensional integral in \eqref{survival_joint_marginal} is reduced to a set of double integrals. Comparing this quantity with the original contributing log-likelihood in \eqref{survival_joint_marginal}, we can see weights $1/(n_i-1)$ equate their first two terms.

Given a dataset of $m$ independent clusters $(O_1,\ldots,O_m)$, we maximize the following empirical composite contributing marginal log-likelihood:
\begin{equation}\label{composite_likelihood}
\mathbb{P}_mclog(O;\beta,\rho,\Lambda):=\frac{1}{m}\overset{m}{\underset{i=1}{\sum}}clog(O_i;\beta,\rho,\Lambda) \ .
\end{equation}

An ordinary maximum likelihood estimator, where $\hat{\Lambda}(t)$ is an absolutely continuous function, does not exist for \eqref{composite_likelihood} due to the infinite-dimensional parameter $\Lambda(t)$. Therefore we only consider $\hat{\Lambda}(t)$ to be c$\grave{a}$dl$\grave{a}$g and redefine its derivative as
\[
\hat{\lambda}(t):= \hat{\Lambda}(t)-\hat{\Lambda}(t-)\ .
\]
It is not difficult to see that in order to maximize \eqref{composite_likelihood}, $\hat{\Lambda}$ shall be a non-decreasing step c$\grave{a}$dl$\grave{a}$g function that only jump at observed failure time points.

To be specific, given a dataset of size $m$, we maximize \eqref{composite_likelihood} over the parameter space:
\begin{eqnarray}
\Theta&:=&\mathcal{B}\times\mathcal{R}\times\mathcal{L}\ ,\label{survival_parameter_space}\\
\mathcal{L}&:=&\{\Lambda(\cdot):\text{ a non-decreasing step c$\grave{a}$dl$\grave{a}$g function in }[0,\tau]\text{ with jumps at observed failure time points}\nonumber\\&&\text{ and }\Lambda(0)=0\}\ . \nonumber
\end{eqnarray}
The resulting non-parametric maximum composite likelihood estimate (NPMCLE) is denoted as $(\hat{\beta}_m,\hat{\rho}_m,\hat{\Lambda}_m)$.

Direct maximization of \eqref{composite_likelihood} is computationally challenging due to unobservable frailties. By treating frailties as missing data, we used a generalized version of the EM algorithm for the composite likelihood, similar to \citeasnoun{ref110}. 
However, the EM algorithm cannot directly estimate the correlation parameter and thus our whole algorithm is a hybrid: we maximize \eqref{composite_likelihood} over $(\beta,\Lambda)$ with a fixed $\rho$ by the composite likelihood EM algorithm, then we directly maximize \eqref{composite_likelihood} over $\rho$ with a fixed $(\beta,\Lambda)$, and iterate between the two steps until convergence.

\subsection{EM algorithm for NPMCLE of $(\beta,\Lambda)$}
Treating frailties as missing data, for observation $O_i$, the composite contributing complete log-likelihood can be written as
\begin{scriptsize}
\begin{eqnarray}
&&\overset{n_i}{\underset{j=1}{\sum}}{\underset{k<j}{\sum}}\frac{1}{n_i-1}\left(\Delta_{ij}\left\{\mathrm{log}\lambda(y_{ij})+\mathrm{log}(w_{ij})+Z_{ij}^T\beta\right\} -e^{Z_{ij}^T\beta}\Lambda(y_{ij})w_{ij}+\Delta_{ik}\left\{\mathrm{log}\lambda(y_{ik})+\mathrm{log}(w_{ik})+Z_{ik}^T\beta\right\} -e^{Z_{ik}^T\beta}\Lambda(y_{ik})w_{ik}\right)\nonumber\\=&&
\overset{n_i}{\underset{j=1}{\sum}}\left(\Delta_{ij}\left\{\mathrm{log}\lambda(y_{ij})+\mathrm{log}(w_{ij})+Z_{ij}^T\beta\right\} -e^{Z_{ij}^T\beta}\Lambda(y_{ij})w_{ij}\right)\label{m_step}
\ .
\end{eqnarray}
\end{scriptsize}
Again, we can see weights $1/(n_i-1)$ equate the composite contributing complete log-likelihood to the full contributing complete log-likelihood.
\begin{enumerate}
\item E-step.\\
According to \citeasnoun{ref110}, expectations of every pairwise contributing complete log-likelihood, conditioning on clustered observations, are
\begin{scriptsize}
\[
\mathrm{E}\left\{\left(\Delta_{ij}\left\{\mathrm{log}\lambda(y_{ij})+\mathrm{log}(w_{ij})+Z_{ij}^T\beta\right\} -e^{Z_{ij}^T\beta}\Lambda(y_{ij})w_{ij}+\Delta_{ik}\left\{\mathrm{log}\lambda(y_{ik})+\mathrm{log}(w_{ik})+Z_{ik}^T\beta \right\}-e^{Z_{ik}^T\beta}\Lambda(y_{ik})w_{ik}\right)\mid X_{ij},X_{ik}\right\}\ .
\]
\end{scriptsize}
One intuitive explanation is that we are transforming $O_i$'s into $O_i^*$'s: every new observation in $O_i^*$ is a pair of original observations in the form of $(X_{ij},X_{ik})$, $j\neq k$. By composite likelihood, these new observations are treated as if they are independent of each other.

Since only $w_{ij}$ and $w_{ik}$ are involved with the parameters of interest, we only need to derive:
\[
\mathrm{E}\left\{w_{ij}\mid X_{ij},X_{ik};\beta,\rho,\Lambda\right\},\quad \mathrm{E}\left\{w_{ik}\mid X_{ij},X_{ik};\beta,\rho,\Lambda\right\}
\] 
for every pair $(j,k)$ in cluster $i$. To be specific, we consider four cases:
\begin{enumerate}
\item $(\Delta_{ij},\Delta_{ik})=(1,1)$
\begin{eqnarray*}
&&\mathrm{E}(w_{ij}\mid Y_{ij},Z_{ij},\Delta_{ij}=1,Y_{ik},Z_{ik},\Delta_{ik}=1;\beta,\rho,\Lambda)\\&=&
\frac{2\left((1-\rho_{jk})u(X_{ik};\beta,\Lambda)+1\right)}{1+u(X_{ij};\beta,\Lambda)+u(X_{ik};\beta,\Lambda)+(1-\rho_{jk})u(X_{ij};\beta,\Lambda)u(X_{ik};\beta,\Lambda)}\\&& \times\frac{\frac{3\left((1-\rho_{jk})u(X_{ik};\beta,\Lambda)+1\right)\left((1-\rho_{jk})u(X_{ij};\beta,\Lambda)+1\right)}{1+u(X_{ij};\beta,\Lambda)+u(X_{ik};\beta,\Lambda)+(1-\rho_{jk})u(X_{ij};\beta,\Lambda)u(X_{ik};\beta,\Lambda)}-2(1-\rho_{jk})}{\frac{2\left((1-\rho_{jk})u(X_{ik};\beta,\Lambda)+1\right)\left((1-\rho_{jk})u(X_{ij};\beta,\Lambda)+1\right)}{1+u(X_{ij};\beta,\Lambda)+u(X_{ik};\beta,\Lambda)+(1-\rho_{jk})u(X_{ij};\beta,\Lambda)u(X_{ik};\beta,\Lambda)}-(1-\rho_{jk})}\ ;
\end{eqnarray*}
\item $(\Delta_{ij},\Delta_{ik})=(1,0)$
\begin{eqnarray*}
&&\mathrm{E}(w_{ij}\mid Y_{ij},Z_{ij},\Delta_{ij}=1,Y_{ik},Z_{ik},\Delta_{ik}=0;\beta,\rho,\Lambda)\\=&&\frac{2(1+(1-\rho_{jk})u(X_{ik};\beta,\Lambda))}{1+u(X_{ij};\beta,\Lambda)+u(X_{ik};\beta,\Lambda)+(1-\rho_{jk})u(X_{ij};\beta,\Lambda)u(X_{ik};\beta,\Lambda)}\ ;
\end{eqnarray*}
\item $(\Delta_{ij},\Delta_{ik})=(0,1)$
\begin{scriptsize}
\begin{eqnarray*}
&&\mathrm{E}(w_{ij}\mid Y_{ij},Z_{ij},\Delta_{ij}=0,Y_{ik},Z_{ik},\Delta_{ik}=1;\beta,\rho,\Lambda)\\=&&\frac{2(1+(1-\rho_{jk})u(X_{ik};\beta,\Lambda))}{1+u(X_{ij};\beta,\Lambda)+u(X_{ik};\beta,\Lambda)+(1-\rho_{jk})u(X_{ij};\beta,\Lambda)u(X_{ik};\beta,\Lambda)} -\frac{1-\rho_{jk}}{1+(1-\rho_{jk})u(X_{ij};\beta,\Lambda)}\ ;
\end{eqnarray*}
\end{scriptsize}
\item $(\Delta_{ij},\Delta_{ik})=(0,0)$
\begin{eqnarray*}
&&\mathrm{E}(w_{ij}\mid Y_{ij},Z_{ij},\Delta_{ij}=0,Y_{ik},Z_{ik},\Delta_{ik}=0;\beta,\rho,\Lambda)\\=&&\frac{1+(1-\rho_{jk})u(X_{ik};\beta,\Lambda)}{1+u(X_{ij};\beta,\Lambda)+u(X_{ik};\beta,\Lambda)+(1-\rho_{jk})u(X_{ij};\beta,\Lambda)u(X_{ik};\beta,\Lambda)}\ .
\end{eqnarray*}
\end{enumerate}
Denote $w^{(l)}_{ij}$ as the average of frailty conditional expectations from the $l^{th}$ iteration E-step; i.e.
\[
w^{(l)}_{ij}={\underset{k\neq j}{\sum}}\frac{1}{n_i-1}
\mathrm{E}\left\{w_{ij}\mid X_{ij},X_{ik};\hat{\beta}^{(l-1)},\rho,\hat{\Lambda}^{(l-1)}\right\}\ ,
\]
where $(\hat{\beta}^{(l-1)},\hat{\Lambda}^{(l-1)})$ are estimates from the $(l-1)^{th}$ iteration's M-step and $\rho$ is some fixed value plugged into the EM algorithm.
\item M-step.\\ 
Since the complete composite log-likelihood is exactly the complete joint log-likelihood, M-step inference is quite straightforward.
 
Plugging the imputed $w_{ij}$'s back into \eqref{m_step}, the composite contributing complete log-likelihood becomes
\begin{small}
\begin{equation}\label{m_step2}
\overset{n_i}{\underset{j=1}{\sum}}\left\{\Delta_{ij}\left[\mathrm{log}\lambda(y_{ij}) + Z_{ij}^T\beta\right]- \left[{\underset{k\neq j}{\sum}}\frac{1}{n_i-1}
\mathrm{E}\left\{w_{ij}\mid X_{ij},X_{ik}\right\}
\right]e^{Z_{ij}^T\beta}\Lambda(y_{ij})
+{\underset{k\neq j}{\sum}}\frac{1}{n_i-1}\mathrm{E} \left[\mathrm{log}(w_{ij})\mid X_{ij},X_{ik}\right]\right\}\ .
\end{equation}
\end{small}
In the $l^{th}$ iteration's M-step, we update our estimates by $(\hat{\beta}^{(l)},\hat{\Lambda}^{(l)})$, where $\hat{\beta}^{(l)}$ solves
\begin{eqnarray}
&&\underset{i=1}{\overset{m}{\sum}}\underset{j=1}{\overset{n_i}{\sum}}\Delta_{ij}\left(Z_{ij}-\frac{\underset{k=1}{\overset{m}{\sum}}\underset{l=1}{\overset{n_k}{\sum}}
Z_{kl}w_{kl}^{(l)}\mathrm{exp}(Z_{kl}^T\beta)1(Y_{kl}\geq Y_{ij})}{\underset{k=1}{\overset{m}{\sum}}\underset{l=1}{\overset{n_k}{\sum}}w_{kl}^{(l)}\mathrm{exp}(Z_{kl}^T\beta)1(Y_{kl}\geq Y_{ij})}\right)=0\ ,\label{survival_beta_EE}\\  \text{and}&&\hat{\Lambda}^{(l)}(t)={\underset{s\leq t}{\sum}}\frac{\underset{i=1}{\overset{m}{\sum}}\underset{j=1}{\overset{n_i}{\sum}}1(Y_{ij}=s)\Delta_{ij}}{\underset{i=1}{\overset{m}{\sum}}\underset{j=1}{\overset{n_i}{\sum}}w_{ij}^{(l)}\mathrm{exp}(Z_{ij}^T\hat{\beta}^{(l)})1(Y_{ij}\geq s)}\ .\nonumber
\end{eqnarray}
\end{enumerate}
Note that \eqref{survival_beta_EE} is equivalent to a partial score equation offset by imputed $w_{ij}$'s from the E-step and the estimate of $\Lambda(t)$ is a Breslow-type estimator.

\subsection{Estimating $\rho$ from \eqref{composite_likelihood}}
For some fixed value $(\beta_1,\Lambda_1)$ , we maximize \eqref{composite_likelihood} over $\rho$ directly; i.e., we solve for $\rho$ from
\begin{scriptsize}
\begin{eqnarray*}
\underset{i=1}{\overset{m}{\sum}}\frac{1}{n_i-1}
{\underset{j<k}{\sum}}\frac{\partial \rho_{jk}}{\partial \rho}&&\left( \frac{2(\rho_{jk}-1)\Delta_{ij}\Delta_{ik}u(X_{ij};\beta_1,\Lambda_1)u(X_{ik};\beta_1,\Lambda_1)-\Delta_{ij}u(X_{ik};\beta_1,\Lambda_1)-\Delta_{ik}u(X_{ij};\beta_1,\Lambda_1)+\Delta_{ij}\Delta_{ik}}{w(X_{ij},X_{ik};\beta_1,\rho,\Lambda_1)}\right.\\&&\quad\left.+(1+\Delta_{ij}+\Delta_{ik})\frac{u(X_{ij};\beta_1,\Lambda_1)u(X_{ik};\beta_1,\Lambda_1)}{v(X_{ij},X_{ik};\beta_1,\rho,\Lambda_1)}\right)=0\ .
\end{eqnarray*}
\end{scriptsize}

\section{Asymptotic Theorems}
Before we present the results, we list the additional technical assumptions for the theoretical results of the NPMCLE. Hereafter, $\tau<\infty$ denotes the endpoint time of the study.\\
C7. There exists some strictly positive constant $a_0$ such that 
\[
\mathrm{pr}(C_{ij}\geq\tau\mid Z_{ij})=\mathrm{pr}(C_{ij}=\tau\mid Z_{ij})\geq a_0\quad\text{  a.s.}\ ;
\]
C8. The covariate $Z$ is bounded.

\begin{lemma}
The true parameter is identifiable from the composite contributing marginal likelihood. Furthermore, the composite Fisher information matrix, which is the average of Fisher information matrices from all pairwise observations, is non-singular along any one-dimensional sub-model.
\end{lemma}

\begin{theorem}
NPMCLE $(\hat{\beta}_m,\hat{\rho}_m,\hat{\Lambda}_m)$ will converge uniformly to the true value $(\beta_0,\rho_0,\Lambda_0)$ as the number of independent clusters $m$ goes to infinity, in the metric space $\mathbb{R}^{p_1}\times \mathbb{R}^{p_2}\times l^\infty[0,\tau]$, where $l^\infty[0,\tau]$ is the linear space consisting of all the bounded functions in $[0,\tau ]$ and is equipped with the total variation norm $||\cdot||_V$, defined as the maximum between the sup norm and the total variation of a function.
\end{theorem}

\begin{theorem}
$\sqrt{m}(\hat{\beta}_m^T-\beta_0^T,\hat{\rho}_m^T-\rho_0^T,\hat{\Lambda}_m-\Lambda_0)^T$ weakly converges to a zero-mean Gaussian process in the same metric space as in Theorem 2.
\end{theorem}

Given a dataset of $m$ independent clusters, in the following we discuss how to estimate the asymptotic standard error of NPMCLE from this dataset. For the empirical composite contributing marginal log-likelihood in \eqref{composite_likelihood}, we can get its Hessian matrix by taking its second derivative at $(\beta,\rho,\Lambda(t_{(1)}),\ldots,\Lambda(t_{(Q)}))$, where $t_{(q)}$'s are ordered observed failure event times. We denote this matrix by $H_m$. The empirical composite score function $S_m:=\sum_{i=1}^{m}S_{m,i}/m$ is derived by taking the first derivative of \eqref{composite_likelihood} at $(\beta,\rho,\Lambda(t_{(1)}),\ldots,\Lambda(t_{(Q)}))$, and we can estimate the covariance of $\sqrt{m}S_m$ by $J_m:=\sum_{i=1}^{m}S_{m,i}S_{m,i}^T/m$. For some arbitrary $h=(h_1,h_2,h_3)$ in which $(h_1,h_2)\in \mathbb{R}^{p_1}\times \mathbb{R}^{p_2}$ and $h_3$ is a bounded function on $[0,\tau]$, we denote $h_m$ as the vector comprising of $h_1$, $h_2$ and $h_3(Y_{ij})$ for which $\Delta_{ij}=1$, $i=1,\ldots,m$ and $j=1,\ldots,n_i$, the following theorem guides us to find an asymptotically consistent estimate of the NPMCLE covariance matrix.
\begin{theorem}
Let $V(h_1,h_2,h_3)$ be the asymptotic covariance matrix of 
\[
\sqrt{m}\left\{h_1^T(\hat{\beta}_m-\beta_0)+h_2^T(\hat{\rho}_m-\rho_0)+\int^\tau_0h_3(s)d[\hat{\Lambda}_m-\Lambda_0](s)\right\}\ .
\]
Then $h_m^TH_m^{-1}J_mH_m^{-1}h_m{\overset{p.}{\longrightarrow}}V(h_1,h_2,h_3)$ uniformly for $(h_1,h_2,h_3)$ such that 
\[
||h_1||\leq 1,\quad ||h_2||\leq 1,\quad ||h_3||_V\leq 1\ .
\]
\end{theorem}
Outlines of the proofs are provided in Appendix A.

To estimate the covariance matrix of $(\hat{\beta}_m,\hat{\rho}_m)$, we can set
\begin{equation}\label{CI.deriv}
h_m^T=\left(\ba{cccc}I& 0_{p_1+p_2} & \cdots& 0_{p_1+p_2} \ea\right)\ ,
\end{equation}
in which $I$ is a $(p_1 +p_2)\times (p_1 +p_2)$ identity matrix, and $0_{p_1+p_2}$ is a zero column vector of length $p_1 +p_2$ and $Q$ such zero vectors are put into $h_m$.

\section{Numerical Studies}
\subsection{Simulations}
We conducted simulations to study the finite sample performance of the proposed hybrid EM algorithm under different censoring rates and correlation structures. Four simulation settings were considered, each based upon 1000 Monte Carlo samples, and every sample contained 200 independent clusters, and cluster sizes varied from $5$ to $7$ with equal probabilities.

We considered two covariates: $Z_1$ is normally distributed with mean $0$ and standard deviation $0.5$, and $Z_2=0.2\times Z_1+Z_0-0.3$, in which $Z_0$ is a Bernoulli variable with mean 0.3. Given $W_{ij}$ and $Z_{ij}$, failure time $T_{ij}$ was generated by:
\[
S(T_{ij}=t\mid W_{ij}=w_{ij},Z_{ij}=z_{ij})=\mathrm{exp}\{-w_{ij}\Lambda_0(t)\,\mathrm{exp}\, (1.2\times z_{ij1} + 2.5\times z_{ij2})\}\ .
\]
The censoring time was the minimum between 10 and an exponential random variable. This exponential random variable is identically and independently distributed (i.i.d.) across observations. Different means of exponential distributions were chosen to generate different censoring rates.

In the first two settings, we considered individual frailties being exchangeably correlated. In the first setting, censoring times were the minimum of an exponential distribution with mean 3.64 and 10, resulting in a censoring rate around 40\%. In the second setting, we adopted an exponential distribution with mean 0.59 for censoring time generation, resulting in a censoring rate around 75\%. Results for finite-dimensional parameter estimates were listed in Table 1. The standard error estimates were based upon Theorem 4 and \eqref{CI.deriv}.

Autoregressive correlation structures have been widely used to model longitudinal data, such as in \citeasnoun{ref106}, and it is also widely used in spatial data analysis, as discussed by \citeasnoun{ref111} and \citeasnoun{ref76}. We found it is also suitable for modeling correlations induced by common genetic factors in some family studies. For example, 100\% of genetic material is shared by monozygotic twins, 50\% is shared by parent-offspring pairs, 25\% is shared by grandparent-offspring pairs, etc. In the last two settings, the same pair of exponential distributions were applied for censoring generation. Results for finite-dimensional parameters were listed in Table 2. The standard error estimates were based upon Theorem 4 and \eqref{CI.deriv}.

When the censoring rate is lower, under both correlation structures, the biases are lower and the standard errors are smaller. Under auto-regressive correlation structure of order one, our estimates are slightly more biased for $\rho$ than the case of exchangeable correlation structure. The estimation performance for $\beta$ are similar across different correlation structure settings. The empirical coverage of 95\% confidence intervals are close to the nominal coverage rate.

\begin{table}[h]
\caption{\sl{Simulation results of estimating $(\beta_0,\beta_1,\rho)$ in different scenarios. $\rho_0$ is the true correlation parameter of frailties, under an exchangeable correlation structure. Bias represents the empirical bias. SEE represents the averaged model-based standard error estimates. SSE represents the Monte Carlo standard error, MSE is the summation of squared SSE and squared bias. Empirical coverage probabilities for 95\% confidence intervals are presented.\label{T1}}}
\begin{small}
\begin{tabular}{p{1.4cm}|
p{0.6cm}|p{0.7cm}p{0.7cm}|p{0.7cm}p{0.7cm}|p{0.7cm}p{0.7cm}|p{0.7cm}p{0.7cm}|p{1cm}p{1.5cm}|p{0.5cm}}
&$\rho_0$&\multicolumn{2}{c|}{Bias$\times 10^{3}$}&\multicolumn{2}{c|}{SEE$\times 10^{3}$}&\multicolumn{2}{c|}{SSE$\times 10^{3}$}&\multicolumn{2}{c|}{MSE$\times 10^{3}$}&\multicolumn{2}{c|}{95\% C.I. coverage rate}&Bias$\times 10^{3}$\\&
&$\hat{\beta}_0$&$\hat{\beta}_1$&$\hat{\beta}_0$&$\hat{\beta}_1$&
$\hat{\beta}_0$&$\hat{\beta}_1$&$\hat{\beta}_0$&$\hat{\beta}_1$&
$\hat{\beta}_0$&$\hat{\beta}_1$&$\hat{\rho}$\\ \hline
\multirow{5}{1.2cm}{Censoring rate 40\%}&0.1&-3&-4&90&136&94&139&9&19&94\%&94\%&2\\
&0.3&-0.7&-4&90&135&92&133&8&18&93\%&95\%&-5\\
&0.5&5&-3&89&135&89&133&8&18&95\%&95\%&-5\\
&0.7&2&0.2&87&133&86&134&7&18&95\%&95\%&-3\\
&0.9&0.6&-3&84&130&86&133&7&18&94\%&94\%&-4\\
\hline
\multirow{5}{1.2cm}{Censoring rate 75\%}&0.1&7&-3&121&172&128&174&16&30&93\%&96\%&13\\
&0.3&3&0.2&121&172&126&174&16&30&95\%&95\%&-10\\
&0.5&2&0.6&119&170&121&175&15&31&94\%&94\%&-13\\
&0.7&-2&2&119&170&120&172&14&30&96\%&95\%&-8\\
&0.9&2&2&117&169&122&171&15&29&94\%&96\%&-15\\ 
\end{tabular}
\end{small}
\end{table}

\begin{table}[h]
\caption{\sl{Simulation results of estimating $(\beta_0,\beta_1,\rho)$ in different scenarios. $\rho_0$ is the true correlation parameter of frailties, under an AR(1) correlation structure, mimicking a spatial study. Bias represents the empirical bias. SEE represents the averaged model-based standard error estimates. SSE represents the Monte Carlo standard error, MSE is the summation of squared SSE and squared Bias. Empirical coverage probabilities for 95\% confidence intervals are presented.\label{T2}}}
\begin{small}
\begin{tabular}{p{1.4cm}|
p{0.6cm}|p{0.7cm}p{0.7cm}|p{0.7cm}p{0.7cm}|p{0.7cm}p{0.7cm}|p{0.7cm}p{0.7cm}|p{1cm}p{1.5cm}|p{0.5cm}}
&$\rho_0$&\multicolumn{2}{c|}{Bias$\times 10^{3}$}&\multicolumn{2}{c|}{SEE$\times 10^{3}$}&\multicolumn{2}{c|}{SSE$\times 10^{3}$}&\multicolumn{2}{c|}{MSE$\times 10^{3}$}&\multicolumn{2}{c|}{95\% C.I. coverage rate}&Bias$\times 10^{3}$\\&
&$\hat{\beta}_0$&$\hat{\beta}_1$&$\hat{\beta}_0$&$\hat{\beta}_1$&
$\hat{\beta}_0$&$\hat{\beta}_1$&$\hat{\beta}_0$&$\hat{\beta}_1$&
$\hat{\beta}_0$&$\hat{\beta}_1$&$\hat{\rho}$\\ \hline
\multirow{5}{1.2cm}{Censoring rate 40\%}& 0.1&0.5&-2&90&136&91&139&8&19&95\%&95\%&7\\
&0.3&2&-3&90&136&90&138&8&19&94\%&94\%&-5\\
&0.5&-2&-5&89&135&90&133&8&18&94\%&95\%&-7\\
&0.7&0.6&-2&88&134&89&136&8&19&94\%&94\%&-6\\
&0.9&2&-0.4&85&131&86&132&7&17&96\%&94\%&-3\\ \hline 
\multirow{5}{1.2cm}{Censoring rate 75\%}&0.1&0.9&1&121&172&121&178&15&32&95\%&94\%&34\\
&0.3&-3&-2&120&172&122&180&15&32&95\%&95\%&-12\\
&0.5&-5&1&120&172&121&175&15&31&94\%&94\%&-31\\
&0.7&-6&-6&120&171&124&172&15&30&94\%&95\%&-21\\
&0.9&-1&-2&118&170&120&173&14&30&94\%&95\%&-10\\ 
\end{tabular}
\end{small}
\end{table}

\subsection{Real Data Analysis}
In the Rats dataset from \citeasnoun{ref98}, three rats were chosen from each of 100 litters, one of which was treated with a drug while the other two served as controls. All mice were followed for tumor incidence in 2 years. A subject is censored if died from other causes. 

We find our model is particularly useful under this design where in each litter, some members get the treatment while the others serve as controls. Conditioning on unobservable environmental or genetic factors, hazard ratio between individuals in treatment and control groups is constantly proportional to each other over time. However, environmental or genetic factors could also affect survival distribution so marginally, this difference in hazard rate due to treatment is finally worn out as time passes. Thus it is reasonable to analyse the Rats dataset with our model. We ran our model with treatment indicator as the sole covariate, assuming an exchangeable correlation structure among rats from the same litter. Censoring rate of this study is approximately 75\% .

We estimated the conditional hazard rate ratio comparing the treatment group to the control group was 2.56 (1.30, 5.02), which is also the marginal log failure odds ratio. Deterioration or tumorigenic effect of treatment on rat survival is statistically significant. We estimated the correlation between frailties to be 0.75. Correlation between individual frailties is high, indicating that there is strong litter effect such as common genetic factors. 

\section{Concluding remarks}

In this paper we introduce a marginalizable individual frailty model for analysing clustered right-censored data. A multivariate exponential distribution for individual frailties is proposed, which yields a marginal proportional odds model that has a population-level interpretation, and the model also allows a flexible correlation structure among observations. Unlike most marginal models, our model guarantees that parameters to be estimated always correspond to some real parameters from a certain distribution.

As for model inference, we maximize a composite contributing marginal log-likelihood. We did not choose to maximize the joint contributing marginal log-likelihood, due to computation complexity. We neither use the penalized log-likelihood, because the density of a multivariate exponential random vector is intractable and the density is needed to specify a penalizion term. Our estimation efficiency is not optimal among all asymptotically linear estimators. However, by only specifying pairwise joint distribution, we gain some level of robustness in return, as pointed out by \citeasnoun{ref61}.

The marginalization property is based on a standard exponential distributed frailty assumption, which can be viewed as a special case of the Gamma frailty assumption that has been well studied by \citeasnoun{ref36} and \citeasnoun{ref38}, etc. We may also consider Gamma distributed frailties; however, marginally it is no longer a proportional odds model. Suppose every frailty $W_{ij}$ is under Gamma distribution with mean one and unknown variance $1/\gamma$, with density
\[
f_\gamma(w_{ij})=\frac{\gamma^\gamma}{\Gamma(\gamma)}w_{ij}^{\gamma -1}e^{-\gamma w_{ij}} \ ,
\]
Integrating over $W_{ij}$ gives the marginal survival probability
\[
S(t\mid Z_{ij})=\left(\frac{1}{1+\Lambda_0(t)e^{Z_{ij}^T\beta-log \gamma}}\right)^\gamma \ .
\]
Thus only at $\gamma=1$, $\beta$ has a marginal proportional failure odds interpretation. Besides, this generalization also ends up with an identifiability problem under an individual frailty model. The variance parameter $\gamma$ is weakly identifiable when the correlation level is small. When correlation level is the highest possible, we obtain a shared frailty model such that $\gamma$ can be estimated routinely. But when correlation level is zero, we get independent observations and $\gamma$ is the power of the survival function, which is quite hard to estimate in general. \citeasnoun{ref22} discussed a quite similar problem in binary data. They came to the same conclusion by showing in Monte Carlo simulations, the condition number, which is the ratio between the maximum and the minimum eigenvalues of Fisher information matrix, is extremely large when the correlation level is low. A large condition number indicates there is little information from the data on for some parameter(s).

Combining the above discussion and the discussion on frailty dispersion, exponential distributed frailties should not be considered as a limitation, since
\begin{enumerate}
\item a marginal proportional odds model interpretation is often desirable in practice;
\item Gumbel distributed random intercepts have physical interpretations and is reasonable when we believe there are many unobservable effects and the maximum dominates the others on affecting the outcome;
\item flexible correlation structures can be imposed on individual frailties; which is proper and beneficial for datasets with complicated correlation structures;
\item The inference of marginal covariate effects is un-affected when frailty distribution is covariate dependent, as discussed in Section 2$\cdot$4.
\end{enumerate}

\newpage
\appendix

\section*{Appendix A}
In this appendix, we outline the proofs of the lemma and theorems in Section 4. We introduce some notations to begin with. Let $\mathbb{P}_m$ be the empirical measure of $m$ i.i.d. cluster observations: $O_1,\ldots,O_m$; denote $P_0$ as the expectation of cluster observations. That is, for any measurable function $g(O)$, we define
\[
\mathbb{P}_m[g(O)]=\frac{1}{m}\underset{i=1}{\overset{m}{\sum}}g(O_i);\quad
P_0[g(O)]=\mathrm{E}[g(O)]\ .
\]
Even though different clusters may contain different numbers of observations, we can still view joint observations from each cluster as i.i.d. samples.

We can regard each cluster in theory contains infinite subjects and their quantities can be denoted by a set of stochastic processes $(Z(\cdot),T(\cdot),C(\cdot),W(\cdot))$: $\cdot$ varies with different subjects. The data we observe from one cluster is a deterministic projection of $(Z(\cdot),T(\cdot),C(\cdot),W(\cdot))$. If we assume the stochastic process sets $(Z(\cdot),T(\cdot),C(\cdot),W(\cdot))$ are i.i.d. and projection procedures are also i.i.d., we can conclude data from different clusters are i.i.d.
\subsection*{\textbf{A.1 Lemma 1}}
\begin{proof}
First, we would like to show when the contributing marginal likelihood for any pair of dependent observations, $(X_j,X_k)$, $j\neq k$ under two sets of parameters $(\Lambda_0,\beta_0,\rho_0)$ and $(\Lambda_1,\beta_1,\rho_1)$, is identical, then
\[
(\beta_1,\rho_1,\Lambda_1)=(\beta_0,\rho_0,\Lambda_0)\ .
\]
Define $t_1^*=\inf\{t:\Lambda_1(t)>0\}$. If $t^*_1 >0 $, by condition C4, it is possible to observe $0 < Y_j<t^*_1\leq Y_k$ and the pairwise contributing marginal likelihood will differ. Thus $t^*_1=0$.

Consider the case $\Delta_j=\Delta_k=1$, taking ratio of their pairwise contributing marginal likelihood, whose logarithm is defined in \eqref{composite}, under these two sets of parameters,
\begin{eqnarray}
&&\frac{L(y_j,y_k,z_j,z_k,\delta_j=\delta_k=1;\beta_0,\rho_0,\Lambda_0)}{L(y_j,y_k,z_j,z_k,\delta_j=\delta_k=1;\beta_1,\rho_1,\Lambda_1)}\nonumber\\&=&\frac{(1-\rho_{0jk})^2\Lambda_0(y_j)e^{z_j^T\beta_0}
\Lambda_0(y_k)e^{z_k^T\beta_0}+(1-\rho_{0jk})\Lambda_0(y_j)e^{z_j^T\beta_0}
+(1-\rho_{0jk})\Lambda_0(y_k)e^{z_k^T\beta_0}+1}{(1-\rho_{1jk})^2\Lambda_1(y_j)e^{z_j^T\beta_1}
\Lambda_1(y_k)e^{z_k^T\beta_1}+(1-\rho_{1jk})\Lambda_1(y_j)e^{z_j^T\beta_1}
+(1-\rho_{1jk})\Lambda_1(y_k)e^{z_k^T\beta_1}+1}\nonumber\\&&\cdot
\left(\frac{(1-\rho_{1jk})\Lambda_1(y_j)e^{z_j^T\beta_1}
\Lambda_1(y_k)e^{z_k^T\beta_1}+\Lambda_1(y_j)e^{z_j^T\beta_1}
+\Lambda_1(y_k)e^{z_k^T\beta_1}+1}{(1-\rho_{0jk})\Lambda_0(y_j)e^{z_j^T\beta_0}
\Lambda_0(y_k)e^{z_k^T\beta_0}+\Lambda_0(y_j)e^{z_j^T\beta_0}
+\Lambda_0(y_k)e^{z_k^T\beta_0}+1}\right)^3\nonumber\\&&\cdot
\frac{d\Lambda_0(y_j)d\Lambda_0(y_k)}{d\Lambda_1(y_j)d\Lambda_1(y_k)}exp\left((z_j+z_k)^T(\beta_0-\beta_1)\right)
=1 \label{identifiable}\ .
\end{eqnarray}
Consider two monotone decreasing sequences $\{y_{jq}: q=1,2,\ldots\}$ and  $\{y_{kq}: q=1,2,\ldots\}$ such that 
\[
y_{jq}\downarrow 0,\quad y_{kq}\downarrow 0\quad\text{as }q\rightarrow\infty\ .
\]
We assume \eqref{identifiable} holds a.e. for every pair $(z_{j q},z_{k q})$ which are two random entries of a stochastic process $Z(\cdot)$ satisfying 
\begin{equation}\label{survival_identi_p1}
\mathrm{pr}\left[ C_j\geq Y_{jq},C_k\geq Y_{kq}\mid z_{j q},z_{k q}\right]>0\ .
\end{equation}
As $q\rightarrow\infty$, the collection of the stochastic process $Z(\cdot)$ satisfying \eqref{survival_identi_p1} grows into the whole space, by conditions C1, C2 and C4. Thus
\begin{equation}\label{identi1}
{\underset{q \rightarrow\infty}{lim}}\left(\frac{L(y_{jq},y_{kq},z_{jq},z_{kq},\delta_j=\delta_k=1;\Lambda_0,\beta_0,\rho_0)}{L(y_{jq},y_{kq},z_{jq},z_{kq},\delta_j=\delta_k=1;\Lambda_1,\beta_1,\rho_1)}\right)=\left(
\frac{d\Lambda_0(0)}{d\Lambda_1(0)}\right)^{2}exp\left\{(z_j +z_k)^T(\beta_0-\beta_1)\right\}
=1\ .
\end{equation}
By C6, $\beta_1=\beta_0$ $F_{Z}$-almost surely. Consequently, $d\Lambda_1(0)=d\Lambda_0(0)$ $F_{Z}$-almost surely.

Now we want to show $\Lambda_1(t)=\Lambda_0(t)$ for $\forall t\in[0,\tau]$.

Since each pair of the contributing marginal likelihood is identical at the two sets of parameters, the contributing marginal likelihood for a single observation, which is an integration of the former one, should also be identical:
\[
\frac{\mathrm{pr}(y_j,z_j,\delta_{j}=1;\Lambda_0,\beta_0,\rho_0)}
{\mathrm{pr}(y_j,z_j,\delta_{j}=1;\Lambda_1,\beta_0,\rho_1)}=\frac{d\Lambda_0(y_j)}{\left(\Lambda_0(y_j)e^{z_j^T\beta_0}+1\right)^2}\left/\frac{d\Lambda_1(y_j)}{\left(\Lambda_1(y_j)e^{z_j^T\beta_0}+1\right)^2}=1\right.\ .
\]
Integrating from 0 to $t \in [0,\tau]$, we get 
\[
\frac{1}{\Lambda_0(t)e^{z_j^T\beta_0}+1}=\frac{1}{\Lambda_1(t)e^{z_j^T\beta_0}+1};\quad \forall t\in [0,\tau]\ .
\]
Thus $\Lambda_1(\cdot)=\Lambda_0(\cdot)$ on $[0,\tau]$. It is trivial to show $\rho_0=\rho_1$.

Second, we will connect the above conclusion with the composite Kullback-Leibler Distance.
\begin{eqnarray}
&&P_0\left[clog(O;\beta_0,\rho_0,\Lambda_0)-clog(O;\beta_1,\rho_1,\Lambda_1)\right]\nonumber
\\=&&\mathrm{E}_n\left\{ \frac{1}{n-1}{\underset{j<k}{\sum}}P_0\left[logL(X_j,X_k;\beta_0,\rho_0,\Lambda_0)-logL(X_j, X_k;\beta_1,\rho_1,\Lambda_1)\right]\mid n\right\}\label{surv_temp_explain}
\\ \geq &&-\mathrm{E}_n\left\{ \frac{1}{n-1}{\underset{j<k}{\sum}}log\mathrm{E}_{X}\left[\frac{L(X_j,X_k;\beta_1,\rho_1,\Lambda_1)}{L(X_j,X_k;\beta_0,\rho_0,\Lambda_0)}\mid n\right]\right\}=0\ .\label{surv_temp_explain2}
\end{eqnarray}
$\mathrm{E}_n$'s in \eqref{surv_temp_explain} and \eqref{surv_temp_explain2} take expectation over random cluster size $n$; equality in \eqref{surv_temp_explain2} holds if and only if $(\beta_0,\rho_0,\Lambda_0)=(\beta_1,\rho_1,\Lambda_1)$.

As for the second part of the lemma, we show it by contradiction.

Suppose there exists some one-dimensional sub-model passing through the true parameters, denoted by $(\beta_0+\epsilon h_1,\rho_0+\epsilon h_2,\Lambda_0+\epsilon\int h_3d\Lambda_0)$ and it has zero value of the composite Fisher information. Equivalently, the score function along this path is zero almost surely for any pair of correlated observations. In this sub-model, correlated observations $j$ and $k$ have score function:
\begin{equation}\label{invertible1}
h_1^Tclog_{\beta}(X_j,X_k;\beta_0,\rho_0,\Lambda_0)+h_{2jk}clog_{\rho_{jk}}(X_j,X_k;\beta_0,\rho_0,\Lambda_0)+clog_\Lambda(X_j,X_k;\beta_0,\rho_0,\Lambda_0)\left[\int h_3d\Lambda_0\right] = 0\text{ a.s.}
\end{equation}
where $clog_{\beta}(X_j,X_k;\beta,\rho,\Lambda)$ is the score function for $\beta$, $clog_{\rho_{jk}}(X_j,X_k;\beta,\rho,\Lambda)$ is the score function for $\rho_{jk}$, $h_{2jk}$ is the entry of $h_2$ corresponding to $\rho_{jk}$ and $clog_{\Lambda}(X_j,X_k;\beta,\rho,\Lambda)\left[\int h_3d\Lambda\right]$ is the score function for $\Lambda$ from the sub-model $\Lambda+\epsilon\int h_3d\Lambda$; . 

With straight-forward calculation it reveals
\begin{scriptsize}
\begin{eqnarray*}
&& h_1^Tclog_{ \beta}(X_j,X_k; \beta,\rho,\Lambda)\\=&& h_1^T\left\{\int^\tau_0Z_jdN_j(s)+ \int^\tau_0Z_kdN_k(s)-\int^\tau_0 A(u,X_j,X_k; \beta,\rho,\Lambda)d\Lambda(u)\right\}\ ,
\\&&
h_{2jk}clog_{\rho_{jk}}(X_j,X_k;  \beta,\rho,\Lambda)\\=&&
 h_{2jk}\left\{
\frac{-2(1-\rho_{jk})\Delta_j\Delta_k
e^{Z_j^T \beta}e^{Z_k^T \beta}\Lambda(Y_j)\Lambda(Y_k)
-\Delta_ke^{Z_j^T \beta}\Lambda(Y_j)
-\Delta_je^{Z_k^T \beta}\Lambda(Y_k)+\Delta_j\Delta_k
}{(1-\rho_{jk})^2\Delta_j\Delta_k
e^{Z_j^T \beta}e^{Z_k^T \beta}\Lambda(Y_j)\Lambda(Y_k)
+(1-\rho_{jk})\Delta_ke^{Z_j^T \beta}\Lambda(Y_j)
+(1-\rho_{jk})\Delta_je^{Z_k^T \beta}\Lambda(Y_k)
+1+\Delta_j\Delta_k\rho_{jk}}\right.\\&&\left.+(1+\Delta_j+\Delta_k)\frac{e^{Z_j^T \beta}e^{Z_k^T \beta}\Lambda(Y_j)\Lambda(Y_k)
}{(1-\rho_{jk})
e^{Z_j^T \beta}e^{Z_k^T \beta}\Lambda(Y_j)\Lambda(Y_k)
+e^{Z_j^T \beta}\Lambda(Y_j)
+e^{Z_k^T \beta}\Lambda(Y_k)+1}\right\}\ ,\\&& clog_{\Lambda}( X_j,X_k; \beta,\rho,\Lambda)\left[\int h_3d\Lambda\right]\\=&& 
\left(\int^\tau_0h_3(s)dN_j(s)+\int^\tau_0h_3(s)dN_k(s)\right)-\int^\tau_0 D(u,X_j,X_k; \beta,\rho,\Lambda)h_3(u)d\Lambda(u)\ ,
\end{eqnarray*}
\end{scriptsize}
where we denote
\begin{scriptsize}
\begin{eqnarray*}
u(X_{ij};\beta,\Lambda)&=&\Lambda(Y_{ij})e^{Z_{ij}^T\beta}\ ,\\
v(X_{ij},X_{ik};\beta,\rho,\Lambda)&=&(1-\rho_{jk})u(X_{ij};\beta,\Lambda)u(X_{ik};\beta,\Lambda)+u(X_{ij};\beta,\Lambda)+u(X_{ik};\beta,\Lambda)+1\ ,\\
w(X_{ij},X_{ik};\beta,\rho,\Lambda)&=&\Delta_{ij}\Delta_{ik}(1-\rho_{jk})^2u(X_{ij};\beta,\Lambda)u(X_{ik};\beta,\Lambda)\\&&+\Delta_{ij}(1-\rho_{jk})u(X_{ik};\beta,\Lambda)+\Delta_{ik}(1-\rho_{jk})u(X_{ij};\beta,\Lambda)+1+\Delta_{ij}\Delta_{ik}\rho_{jk}\ ,
\end{eqnarray*}

\begin{eqnarray*}
D(u,X_{ij},X_{ik};\beta,\rho,\Lambda)&:=&(1+\Delta_{ij}+\Delta_{ik})
\frac{e^{Z_{ij}^T\beta}1\{Y_{ij}\geq u\}\left[1+(1-\rho_{jk})e^{Z_{ik}^T\beta}\Lambda(Y_{ik})\right]+
e^{Z_{ik}^T\beta}1\{Y_{ik}\geq u\}\left[1+(1-\rho_{jk})e^{Z_{ij}^T\beta}\Lambda(Y_{ij})\right]
}{v(X_{ij},X_{ik};\beta,\Lambda,\rho)}
\\&&-\frac{\Delta_{ik}(1-\rho_{jk}) e^{Z_{ij}^T\beta}1\{Y_{ij}\geq u\}
\left[
\Delta_{ij}(1-\rho_{jk})e^{Z_{ik}^T\beta}\Lambda(Y_{ik})+1\right]}
{w(X_{ij},X_{ik};\beta,\Lambda,\rho)}
\\&&
-\frac{\Delta_{ij}(1-\rho_{jk}) e^{Z_{ik}^T\beta}1\{Y_{ik}\geq u\}
\left[
\Delta_{ik}(1-\rho_{jk})e^{Z_{ij}^T\beta}\Lambda(Y_{ij})+1\right]
}{w(X_{ij},X_{ik};\beta,\Lambda,\rho)}
\ ,\\
A(u,X_{ij},X_{ik};\beta,\rho,\Lambda)&=&Z_{ij}e^{Z_{ij}^T\beta}1\{Y_{ij}\geq u\}\left\{(1+\Delta_{ij}+\Delta_{ik})
\frac{1+(1-\rho_{jk})e^{Z_{ik}^T\beta}\Lambda(Y_{ik})}{v(X_{ij},X_{ik};\beta,\Lambda,\rho)}
-\frac{\Delta_{ik}(1-\rho_{jk}) 
\left[
\Delta_{ij}(1-\rho_{jk})e^{Z_{ik}^T\beta}\Lambda(Y_{ik})+1\right]}{w(X_{ij},X_{ik};\beta,\Lambda,\rho)}\right\}\\&&+Z_{ik}e^{Z_{ik}^T\beta}1\{Y_{ik}\geq u\}\left\{(1+\Delta_{ij}+\Delta_{ik})
\frac{1+(1-\rho_{jk})e^{Z_{ij}^T\beta}\Lambda(Y_{ij})}{v(X_{ij},X_{ik};\beta,\Lambda,\rho)}
-\frac{\Delta_{ij}(1-\rho_{jk}) 
\left[
\Delta_{ik}(1-\rho_{jk})e^{Z_{ij}^T\beta}\Lambda(Y_{ij})+1\right]}{w(X_{ij},X_{ik};\beta,\Lambda,\rho)}\right\}\ .
\end{eqnarray*}
\end{scriptsize}
Consider the case $\Delta_j=\Delta_k=1$ and two monotone decreasing sequences $\{y_{jq}: q=1,2,\ldots\}$ and  $\{y_{kq}: q=1,2,\ldots\}$ such that 
\[
y_{jq}\downarrow 0\quad\text{as }q\rightarrow\infty,\quad y_{kq}\downarrow 0\quad\text{as }q\rightarrow\infty\ .
\]
Since
\[
{\underset{q\rightarrow\infty}{\text{lim}}}\int^\tau_0 D(u,X_{jq},X_{kq};\beta_0,\rho_0,\Lambda_0)h_3(u) d\Lambda_0(u) \leq { \underset{q\rightarrow\infty}{\text{lim}}} [Y_{jq}\vee Y_{kq}]\cdot M_3\cdot h_3(u) \Lambda_0(Y_{jq}\vee Y_{kq})\rightarrow 0\ ,
\]
taking limit of \eqref{invertible1} in $q$, we get
\begin{equation}\label{invertible5}
h_1^T\left(Z_j+Z_k\right)+h_{2jk}\frac{1}{1+\rho_{jk}}+2h_3(0)=0\ .
\end{equation}
Thus $h_1=0$ $F_{Z}$ a.e., using the same argument from the first part.

Consider another pair of observations $(X_j=(Y_j=\tau,\Delta_j=0,Z_j), \, X_k=(Y_k=\tau,\Delta_k=0,Z_k))$. According to condition C7, we know any $Z(\cdot)$ from the covariate sample space corresponds to such pair of observations with positive probability. 
Then we write \eqref{invertible1} as
\begin{eqnarray}
&&h_{2jk}\frac{e^{Z_j^T\beta_0}e^{Z_k^T\beta_0}\Lambda_0(\tau)\Lambda_0(\tau)
}{(1-\rho_{jk})
e^{Z_j^T\beta_0}e^{Z_k^T\beta_0}\Lambda_0(\tau)\Lambda_0(\tau)
+e^{Z_j^T\beta_0}\Lambda_0(\tau)
+e^{Z_k^T\beta_0}\Lambda_0(\tau)+1}\nonumber\\&&-
\frac{\left[1+(1-\rho_{jk})e^{Z_k^T\beta_0}\Lambda_0(\tau)\right]e^{Z_j^T\beta_0}+
\left[1+(1-\rho_{jk})e^{Z_j^T\beta_0}\Lambda_0(\tau)\right]e^{Z_k^T\beta_0}
}{(1-\rho_{jk})
e^{Z_j^T\beta_0}e^{Z_k^T\beta_0}\Lambda_0(\tau)\Lambda_0(\tau)
+e^{Z_j^T\beta_0}\Lambda_0(\tau)
+e^{Z_k^T\beta_0}\Lambda_0(\tau)+1}
\int^\tau_0 h_3(u)d\Lambda_0(u)=0\ ;\nonumber\\&& \text{i.e. } h_{2jk}e^{Z_j^T\beta_0}e^{Z_k^T\beta_0}\Lambda_0(\tau)\Lambda_0(\tau)
\nonumber\\&&-\left(
\left[1+(1-\rho_{jk})e^{Z_k^T\beta_0}\Lambda_0(\tau)\right]e^{Z_j^T\beta_0}+
\left[1+(1-\rho_{jk})e^{Z_j^T\beta_0}\Lambda_0(\tau)\right]e^{Z_k^T\beta_0}
\right)\int^\tau_0 h_3(u)d\Lambda_0(u)=0\nonumber\ .\\&&\label{invertible2}
\end{eqnarray}
Consider another covariate $Z_j'=Z_j+log2/\beta$ and plug it into \eqref{invertible2}:
\begin{eqnarray*}
&&2h_{2jk}e^{Z_j^T\beta}e^{Z_k^T\beta}\Lambda_0(\tau)\Lambda_0(\tau)
\nonumber\\&&-\left(
2\left[1+(1-\rho_{jk})e^{Z_k^T\beta}\Lambda_0(\tau)\right]e^{Z_j^T\beta}+
\left[1+2(1-\rho_{jk})e^{Z_j^T\beta}\Lambda_0(\tau)\right]e^{Z_k^T\beta}
\right)\int^\tau_0 h_3(u)d\Lambda(u)=0\ .\\ &&\text{If }h_{2jk}\neq 0\ , \text{then } 
1+2(1-\rho_{jk})e^{Z_j^T\beta}\Lambda_0(\tau)=2+2(1-\rho_{jk})e^{Z_j^T\beta}\Lambda_0(\tau)\ .
\end{eqnarray*}
Contradiction is achieved. Thus, we need to have $h_{2jk}=0$ a.e. w.r.t. $F_{Z}$ and thus $h_{2}=0$; consequently, $h_3(0)=0$ by \eqref{invertible5}.

Up to now we can claim that
\begin{equation}\label{invertible3}
\int^\tau_0h_3(u)d\left(N_j(u)+N_k(u)\right) -\int^\tau_0D(u,X_j,X_k;\beta_0,\rho_0,\Lambda_0)h_3(u)d\Lambda_0(u)=0\ ,
\end{equation}
for arbitrary pairs of observations.

Consider the case $\Delta_j=\Delta_k=1$ and one monotone decreasing sequence $\{y_{kq}: q=1,2,\ldots\}$ such that 
\[
y_{kq}\downarrow 0\quad\text{as }q\rightarrow\infty\ .
\]
Let $(Z_{j},Z_{k q})$ be two random entries from some stochastic process $Z(\cdot)$ satisfying 
\begin{equation}\label{survival_identi_p2}
Pr\left[ C_j\geq y_{j},C_k\geq y_{kq}\mid Z_{j},Z_{k q}\right]>0\ .
\end{equation}
As $q\rightarrow\infty$, collection of stochastic process $Z(\cdot)$ satisfying \eqref{survival_identi_p2} grows into a subset of the whole covariate space and we restrict ourself working in this subset.
\begin{eqnarray}
&&h_3(Y_j)-e^{Z_j^T\beta_0}\left(\frac{3}{
e^{Z_j^T\beta_0}\Lambda_0(Y_j)+1}-\frac{1-\rho_{jk}}{(1-\rho_{jk})e^{Z_j^T\beta_0}\Lambda_0(Y_j)+1+\rho_{jk}}\right)\int_0^{Y_j}h_3(u)d\Lambda_0(u)
=0\nonumber\ ;\\&&\text{i.e.}\quad
h_3(Y_j)-C_0\int_0^{Y_j}h_3(u)d\Lambda_0(u)
=0\ .\label{invertible3.1}
\end{eqnarray}
Since $h_3(u)\lambda_0(u)$ is a $L_1(P_0)$ function, then its integral over $u$, $h_3(Y_j)\Lambda_0(Y_j)$ is absolutely continuous on $[0,\tau]$.
Thus, taking derivative of $Y_j$ in \eqref{invertible3.1}, we get
\begin{equation}\label{invertible3.2}
h'_3(Y_j)+C_1h_3(Y_j)\lambda_0(Y_j)=0\quad \forall Y_j \in [0,\tau]\ .
\end{equation}
Solving the differential equation in \eqref{invertible3.2}:
\[
e^{C_1\Lambda_0(Y_j)}h_3(Y_j)=\text{constant.}
\]
Since $h_3(0)=0$ and $Y_j$ can be arbitrary value in $[0,\tau]$ due to C4, we claim $h_3(u)=0$ for $\forall u \in [0,\tau]$.
\end{proof}
\subsection*{\textbf{A.2 Theorem 1}}
Consistency of the NPMCLE can be demonstrated by first showing that $\hat{\Lambda}_m(\tau)$ is uniformly bounded a.s.. 
Then by Helly's selection lemma and the compactness of $\mathcal{B}\times\mathcal{R}$, for every subsequence of NPMCLE $\{\hat{\theta}_n\}=\{(\hat{\beta}_n,\hat{\rho}_n,\hat{\Lambda}_n)\}$, there exists a subsequence $\{\hat{\theta}_{n'}\}$ such that $\hat{\theta}_{n'}\rightarrow \theta^*:=(\beta^*,\rho^*,\Lambda^*)$, which is an inner point of parameter space $\Theta$. This convergence is point-wise but we should strengthen it into uniform convergence. The whole proof will be completed if we can show $\theta^*=\theta_0$. However, we cannot write out $\Lambda^*$ explicitly. Therefore, we switch to an intermediate function sequence $\{\tilde{\Lambda}_{n'}\}$ which converges to $\Lambda_0$ uniformly on $[0,\tau]$. In the following we present several key steps of the proof, following the structure provided by \citeasnoun{ref90}.

\begin{proof}
To show the uniform boundedness of $\hat{\Lambda}_m(\tau)$, we compare the values of the empirical composite contributing marginal log-likelihood values evaluated at the NPMCLE and another set of parameters and show if $\hat{\Lambda}_m(\tau)$ is not uniformly bounded, then the empirical composite contributing marginal log-likelihood under NPMCLE will go to negative infinity as $m\rightarrow\infty$.
\begin{enumerate}
\item Construct a step function $\bar{\Lambda}_m$.\\
We define
\begin{eqnarray}
&&\bar{\Lambda}_m(t)=\frac{1}{m} \overset{m}{\underset{i=1}{\sum}}\overset{n_i}{\underset{j=1}{\sum}}\Delta_{ij}I(Y_{ij}\leq t),\quad t\in [0,\tau]\nonumber\ .\\ \text{Consequently,}&&\bar{\Lambda}_m(t)=O(1)\ ,\quad\quad \Delta\bar{\Lambda}_m(t)=O(1/m)\nonumber\ ,\\
\text{and}&&\mathbb{P}_m clog(X;\beta_0,\rho_0,\bar{\Lambda}_m)=O(1)+\frac{1}{m}\underset{i=1}{\overset{m}{\sum}}\underset{j=1}{\overset{n_i}{\sum}}\Delta_{ij}\mathrm{log}(1/m)\ .\label{bar_simple}
\end{eqnarray}
\item Empirical composite contributing marginal log-likelihood evaluated at the NPMCLE
\begin{eqnarray}
&&\mathbb{P}_{m}\left\{\frac{1}{n-1}{\underset{j<k}{\sum}}l(X_j,X_k;\hat{\beta}_{m},\hat{\rho}_{m},\hat{\Lambda}_{m})\right\}\nonumber\\=
&&O(1)+\frac{1}{m}\underset{i=1}{\overset{m}{\sum}}\underset{j=1}{\overset{n_i}{\sum}}\Delta_{ij}\mathrm{log}\Delta\hat{\Lambda}_{m}(y_{ij})-\frac{1}{m}\underset{i=1}{\overset{m}{\sum}}\underset{j=1}{\overset{n_i}{\sum}}(1+\Delta_{ij})\mathrm{log}\left(\hat{\Lambda}_{m}(y_{ij})+1\right)\ .\label{composite_simple}
\end{eqnarray}
\item Establish contradiction by assuming $\hat{\Lambda}_m(\tau)\rightarrow\infty$.\\
Consider a partition $\tau=s_0>s_1>\ldots>s_N>s_{N+1}=0$ and the difference between \eqref{composite_simple} and \eqref{bar_simple} follows:
\begin{small}
\begin{eqnarray}
&&\eqref{composite_simple} - \eqref{bar_simple}\nonumber\\
&&=\frac{1}{m}\underset{i=1}{\overset{m}{\sum}}\underset{j=1}{\overset{n_i}{\sum}}\Delta_{ij}\mathrm{log}\left( m\Delta\hat{\Lambda}_{m}(Y_{ij})\right)-\frac{1}{m}\underset{i=1}{\overset{m}{\sum}}\underset{j=1}{\overset{n_i}{\sum}} (1+\Delta_{ij})
\mathrm{log}\left(\hat{\Lambda}_{m}(Y_{ij})+1\right)+O(1)\nonumber\\=&&
\underset{q=0}{\overset{N}{\sum}}\frac{1}{m}\underset{i=1}{\overset{m}{\sum}}\underset{j=1}{\overset{n_i}{\sum}}\Delta_{ij}1\left\{Y_{ij}\in[s_{q+1},s_q)\right\}\mathrm{log}\left( m\Delta\hat{\Lambda}_{m}(Y_{ij})\right)
-\frac{1}{m}\underset{i=1}{\overset{m}{\sum}}\underset{j=1}{\overset{n_i}{\sum}} 1\left\{Y_{ij}=\tau\right\}(1+\Delta_{ij})
\mathrm{log}\left(\hat{\Lambda}_{m}(\tau)+1\right)
\nonumber\\&-&\underset{q=0}{\overset{N}{\sum}}\frac{1}{m}\underset{i=1}{\overset{m}{\sum}}\underset{j=1}{\overset{n_i}{\sum}} 1\left\{Y_{ij}\in[s_{q+1},s_q)\right\}(1+\Delta_{ij})
\mathrm{log}\left(\hat{\Lambda}_{m}(Y_{ij})+1\right)+O(1)\nonumber\\ \leq&&
\underset{q=0}{\overset{N}{\sum}}\frac{1}{m}\underset{i=1}{\overset{m}{\sum}}\underset{j=1}{\overset{n_i}{\sum}}1\left\{Y_{ij}\in[s_{q+1},s_q)\right\}\Delta_{ij}\mathrm{log}\left( m\Delta\hat{\Lambda}_{m}(Y_{ij})\right)
-\frac{1}{m}\underset{i=1}{\overset{m}{\sum}}\underset{j=1}{\overset{n_i}{\sum}} 1\left\{Y_{ij}=\tau\right\}(1+\Delta_{ij})
\mathrm{log}\left(\hat{\Lambda}_{m}(\tau)+1\right)
\nonumber\\&-&\underset{q=0}{\overset{N}{\sum}}\frac{1}{m}\underset{i=1}{\overset{m}{\sum}}\underset{j=1}{\overset{n_i}{\sum}} 1\left\{Y_{ij}\in[s_{q+1},s_q)\right\}(1+\Delta_{ij})
\mathrm{log}\left(\hat{\Lambda}_{m}(s_{q+1})+1\right)+O(1)\ .\label{u1}
\end{eqnarray}
\end{small}
Since $log(x)$ is a concave function, by Jensen's Inequality, 
\[
\frac{\frac{1}{m}\underset{i=1}{\overset{m}{\sum}}\underset{j=1}{\overset{n_i}{\sum}}\Delta_{ij}1\left\{Y_{ij}\in[s_{q+1},s_q)\right\}\mathrm{log}\left( m\Delta\hat{\Lambda}_{m}(Y_{ij})\right)}{\frac{1}{m}\underset{i=1}{\overset{m}{\sum}}\underset{j=1}{\overset{n_i}{\sum}}\Delta_{ij}1\left\{Y_{ij}\in[s_{q+1},s_q)\right\}}\leq \mathrm{log}\left( m \frac{\underset{i=1}{\overset{m}{\sum}}\underset{j=1}{\overset{n_i}{\sum}}\Delta_{ij}1\left\{Y_{ij}\in[s_{q+1},s_q)\right\}\Delta\hat{\Lambda}_{m}(Y_{ij})}{\underset{i=1}{\overset{m}{\sum}}\underset{j=1}{\overset{n_i}{\sum}}\Delta_{ij}1\left\{Y_{ij}\in[s_{q+1},s_q)\right\}}\right)\ .
\]
Thus
\begin{small}
\[
\frac{1}{m}\underset{i=1}{\overset{m}{\sum}}\underset{j=1}{\overset{n_i}{\sum}}\Delta_{ij}1\left\{Y_{ij}\in[s_{q+1},s_q)\right\}\mathrm{log}\left( m\Delta\hat{\Lambda}_{m}(Y_{ij})\right) \leq 
O(1) + \frac{1}{m}\underset{i=1}{\overset{m}{\sum}}\underset{j=1}{\overset{n_i}{\sum}}\Delta_{ij}1\left\{Y_{ij}\in[s_{q+1},s_q)\right\}\times  \mathrm{log} \hat{\Lambda}_{m}(s_q)\ .
\]
\end{small}
Then the right side of \eqref{u1} is bounded from above by
\begin{eqnarray}
&&-\underset{q=0}{\overset{N-1}{\sum}}\frac{1}{m}\underset{i=1}{\overset{m}{\sum}}\underset{j=1}{\overset{n_i}{\sum}}\left[
(1+\Delta_{ij})1\left\{Y_{ij}\in[s_{q+1},s_q)\right\}-
\Delta_{ij}1\left\{Y_{ij}\in[s_{q+2},s_{q+1})\right\}\right]
\mathrm{log}\left(\hat{\Lambda}_m(s_{q+1})+1\right)\nonumber\\
&&-\frac{1}{m}\underset{i=1}{\overset{m}{\sum}}\underset{j=1}{\overset{n_i}{\sum}}\left[1\left\{Y_{ij}=\tau\right\}(1+\Delta_{ij})-1\left\{Y_{ij}\in[s_1,\tau)\right\}\Delta_{ij}\right]\mathrm{log}\left(\hat{\Lambda}(\tau)+1\right)\nonumber\\&&
-\frac{1}{m}\underset{i=1}{\overset{m}{\sum}}\underset{j=1}{\overset{n_i}{\sum}}1\{Y_{ij}\in[0,s_N)\}(1+\Delta_{ij})\mathrm{log}(\hat{\Lambda}_m(0)+1)\ .\label{u2}
\end{eqnarray}
We make the partition from $\tau$ to 0. First, we find some $s_1\in[0,\tau)$ such that 
\[
\frac{1}{2}
\mathrm{E}\left\{\underset{j=1}{\overset{n_i}{\sum}}1\left\{Y_{ij}=\tau\right\}\right\}=\frac{1}{2}
\mathrm{E}\left\{\underset{j=1}{\overset{n_i}{\sum}}(1+\Delta_{ij})1\left\{Y_{ij}=\tau\right\}\right\}>
\mathrm{E}\left\{\underset{j=1}{\overset{n_i}{\sum}}\Delta_{ij}1\left\{Y_{ij}\in[s_1,\tau)\right\}\right\}\ .
\]
By conditions C4 and C7, such an $s_1$ exists.

Define a constant $\epsilon\in(0,1)$ such that
\begin{equation}\label{survival_uniform_bounded_epsilon}
\frac{\epsilon}{1-\epsilon}<\frac{\mathrm{E}\left\{\underset{j=1}{\overset{n_i}{\sum}}1\left\{Y_{ij}\in [s_1,s_0)\right\}\right\}}{\mathrm{E}\left\{\underset{j=1}{\overset{n_i}{\sum}}\Delta_{ij}1\left\{Y_{ij}\in [0,s_0)\right\}\right\}}\ .
\end{equation}
If $s_1>0$, we can choose $s_2=0\vee s$ such that $s$ is the minimum value less than $s_1$ satisfying:
\[
(1-\epsilon)\mathrm{E}\left\{\underset{j=1}{\overset{n_i}{\sum}}(1+\Delta_{ij})1\left\{Y_{ij}\in [s_1,s_0)\right\}\right\}\geq 
\mathrm{E}\left\{\underset{j=1}{\overset{n_i}{\sum}}\Delta_{ij}1\left\{Y_{ij}\in [s,s_1)\right\}\right\}\ .
\]
Clearly, $s_2$ exists. The process can continue so that we obtain a sequence $\tau=s_0>s_1>\ldots\geq 0$ such that
\begin{eqnarray*}
\frac{1}{2}
\mathrm{E}\left\{\underset{j=1}{\overset{n_i}{\sum}}(1+\Delta_{ij})1\left\{Y_{ij}=\tau\right\}\right\}&>&
\mathrm{E}\left\{\underset{j=1}{\overset{n_i}{\sum}}\Delta_{ij}1\left\{Y_{ij}\in[s_1,\tau)\right\}\right\}\ ,\\
(1-\epsilon)\mathrm{E}\left\{\underset{j=1}{\overset{n_i}{\sum}}(1+\Delta_{ij})1\left\{Y_{ij}\in [s_q,s_{q-1})\right\}\right\}&\geq& 
\mathrm{E}\left\{\underset{j=1}{\overset{n_i}{\sum}}\Delta_{ij}1\left\{Y_{ij}\in [s_{q+1},s_q)\right\}\right\},\quad q\geq 1\ .
\end{eqnarray*}
\citeasnoun{ref91} have shown, for a quite similar case, that there exists a finite $N$ such that $s_{N+1}=0$. 

If $\{\hat{\Lambda}_m\}$ is not uniformly bounded, then there exists a subsequence $\{\hat{\Lambda}_{m'}\}$ such that $\hat{\Lambda}_{m'}(\tau)\rightarrow\infty$. If this is true, \eqref{u2} will go to negative infinity, contradicting the definition of the NPMCLE.

Therefore, $\hat{\Lambda}_m(\tau)$ is uniformly bounded. 
\end{enumerate}

Let us rewrite the parameter space for NPMCLE inference:
\begin{eqnarray}
\Theta&:=&\mathcal{B}\times\mathcal{R}\times\mathcal{L}\ ,
\label{survival_parameter_space2}\\
\mathcal{L}&:=&\{\Lambda(\cdot):\text{ non-decreasing step c$\grave{a}$dl$\grave{a}$g function in }[0,\tau]\text{ with jumps at observed failure time points}\nonumber\\&&\text{ and }\Lambda(0)=0, \Lambda(\tau)=V_0<\infty\}\ . \nonumber
\end{eqnarray}

Before defining an intermediate term used in this proof, we consider a path through $\hat{\Lambda}_m$, indexed by $\epsilon$, in the direction of $h_t$:
\[
\hat{\Lambda}^\epsilon_m(t)=\int^t_0 (1+\epsilon h_t(s))d\hat{\Lambda}_m(s)\ ,
\]
and take derivative of empirical composite contributing marginal log-likelihood under $(\hat{\beta}_m,\hat{\rho}_m,\hat{\Lambda}^\epsilon_m)$ of $\epsilon$ at $\epsilon=0$, setting $h_t(s)=1\left\{s\leq t\right\}$. Then we can rewrite $\hat{\Lambda}_m$ as
\[
\hat{\Lambda}_m(t)=\int^t_0 \frac{1}{W_m(u;\hat{\beta}_m,\hat{\rho}_m,\hat{\Lambda}_m)}dG_m(u)\ ,
\]
where
\begin{eqnarray*}
W(u,X;\beta,\rho,\Lambda)&=&
\frac{1}{n-1}{\underset{k < j}{\sum}}D(u,X_j,X_k;\beta,\rho,\Lambda)\ ,\\
W_m(u;\beta,\rho,\Lambda)&=&\mathbb{P}_m\left[W(u,X;\beta,\rho,\Lambda)\right],\quad W_0(u;\beta,\rho,\Lambda)=P_0\left[W(u,X;\beta,\rho,\Lambda)\right]\ ,\\
G(t)&=&\overset{n}{\underset{j=1}{\sum}}\Delta_j 1\{Y_j\leq t\}\ ,\\
G_m(t)&=&\mathbb{P}_m\left[\overset{n}{\underset{j=1}{\sum}}\Delta_j 1\{Y_j\leq t\}\right],\quad G_0(t)=P_0\left[\overset{n}{\underset{j=1}{\sum}}\Delta_j 1\{Y_j\leq t\}\right]\ .
\end{eqnarray*}
We define the intermediate term $\tilde{\Lambda}_m$ as
\[
\tilde{\Lambda}_m(t):=\int^t_0\frac{1}{W_m(u;\beta_0,\rho_0,\Lambda_0)}dG_m(u)\ .
\]
It is not hard to verify that $\tilde{\Lambda}_m(t)$ uniformly converges to $\Lambda_0(t)$ by noticing the following two function classes indexed by $t$ and $(u,\theta)$ respectively
\begin{eqnarray*}
\mathcal{F}_1&:=&\left\{f_t(X):=
\int^t_0 g(s)dG(s) :g \text{ is a c$\grave{a}$dl$\grave{a}$g function on  $[0,\tau]$ and } ||g||_V\leq M_1<\infty\right\}\ ,\\
\mathcal{W}&:=&\{W(u,X;\beta,\rho,\Lambda): u\in[0,\tau],(\beta,\rho,\Lambda)\in\Theta\}\ ,
\end{eqnarray*}
are $P_0$-Donsker. Proofs of these two function classes are sketched in Appendix B.

Next we do not apply Helly's Selection Lemma directly to the subsequence $\{\hat{\Lambda}_n\}$. Instead, we define a point-wise converging subsequence of NPMLCE ${n'}$ as
\[
\hat{\beta}_{n'}\rightarrow\beta^*,\quad \hat{\rho}_{n'}\rightarrow\rho^*,\quad W_{n'}(\cdot;\hat{\beta}_{n'},\hat{\rho}_{n'},\hat{\Lambda}_{n'})\rightarrow W^*(\cdot)\text{ pointwise.}
\]
Define 
\[\Lambda^*(t):=\int^t_0\frac{1}{W^*(u)}dG_0(u)\ .\]
By Dominant Convergence Theorem, we can show $\hat{\Lambda}_{n'}$ converges to $\Lambda^*$ uniformly.

By definition of the four sets of parameters (sequences), we can write two quantities $d\hat{\Lambda}_{n'}(t)/d\tilde{\Lambda}_{n'}(t)$ and $d\Lambda^*(t)/d\Lambda_0(t)$. According to Lemma 1 shown by \citeasnoun{ref99}, $d\hat{\Lambda}_{n'}(t)/d\tilde{\Lambda}_{n'}(t)\rightarrow d\Lambda^*(t)/d\Lambda_0(t)$ uniformly.

Since $\{clog(X;\hat{\beta}_{n'},\hat{\rho}_{n'},\hat{\Lambda}_{n'})-clog(X;\beta_0,\rho_0,\tilde{\Lambda}_{n'})\}$ is $P_0$-Glivenko-Cantelli, as shown in Appendix B, we come up with 
\begin{eqnarray*}
&&\mathbb{P}_{n'}\{clog(X;\hat{\beta}_{n'},\hat{\rho}_{n'},\hat{\Lambda}_{n'})-clog(X;\beta_0,\rho_0,\tilde{\Lambda}_{n'})\}\geq 0 \ , \\
\text{implying}&&P_0\{clog(X;\hat{\beta}_{n'},\hat{\rho}_{n'},\hat{\Lambda}_{n'})-clog(X;\beta_0,\rho_0,\tilde{\Lambda}_{n'})\}\geq -o(1)\ .
\end{eqnarray*}
Together with the above proved (uniform) convergence sequences:
\begin{eqnarray*}
&&\hat{\beta}_{n'}\rightarrow\beta^*,\quad \hat{\rho}_{n'}\rightarrow\rho^*,\quad \hat{\Lambda}_{n'}\rightarrow \Lambda^*,\quad d\hat{\Lambda}_{n'}/d\tilde{\Lambda}_{n'}\rightarrow d\Lambda^*/d\Lambda_0\ ;\\ \text{we get}&& P_0\{clog(X;\hat{\beta}_{n'},\hat{\rho}_{n'},\hat{\Lambda}_{n'})-clog(X;\beta_0,\rho_0,\tilde{\Lambda}_{n'})\}\rightarrow P_0\{clog(X;\beta^*,\rho^*,\Lambda^*)-clog(X;\beta_0,\rho_0,\Lambda_0)\}\ .
\end{eqnarray*}
By model identifiability with regards to the composite Kullback-Leibler distance in Lemma 1, 
\[
\beta^*=\beta_0,\quad \rho^*=\rho_0,\quad\Lambda^*=\Lambda_0\ .
\] 
Therefore, consistency is achieved.
\end{proof}

\subsection*{\textbf{A.3 Theorem 2}}
The weak convergence follows from Theorem 3.3.1 in \citeasnoun{book5}, and we need to show that the following conditions for this theorem is satisfied.

Suppose there are two random mappings $\Psi_m$ and $\Psi$, to be defined later, such that $\Psi(\beta_0,\rho_0,\Lambda_0)=0$ 
for some interior point $(\beta_0,\rho_0,\Lambda_0)\in\Theta$, $\Psi_m(\hat{\beta}_m,\hat{\rho}_m,\hat{\Lambda}_m){\overset{P}{\rightarrow}}0$ for some random sequence $(\hat{\beta}_m,\hat{\rho}_m,\hat{\Lambda}_m)\subset \Theta$,
and the followings are true:

\noindent P.1 $(\hat{\beta}_m,\hat{\rho}_m,\hat{\Lambda}_m)$ is consistent for $(\beta_0,\rho_0,\Lambda_0)$;

\noindent P.2 $\sqrt{m}\left(\Psi_m-\Psi\right)(\beta_0,\rho_0,\Lambda_0)$ converges in distribution to a tight random element $Z$;

\noindent P.3 
\begin{eqnarray*}
&&\sqrt{m}\left(\Psi_m-\Psi\right)(\hat{\beta}_m,\hat{\rho}_m,\hat{\Lambda}_m)-
\sqrt{m}\left(\Psi_m-\Psi\right)(\beta_0,\rho_0,\Lambda_0)\\=&&o_p\left(
1+\sqrt{m}||\hat{\beta}_m-\beta_0||+\sqrt{m}||\hat{\rho}_m-\rho_0||+\sqrt{m}||\hat{\Lambda}_m-\Lambda_0||_\infty
\right)\ ;
\end{eqnarray*}

\noindent P.4 $\Psi(\beta,\rho,\Lambda)$ is Fr\'{e}chet differentiable at $(\beta_0,\rho_0,\Lambda_0)$;

\noindent P.5 The derivative of $\Psi(\beta,\rho,\Lambda)$ in $(\beta,\rho,\Lambda)$ at $(\beta_0,\rho_0,\Lambda_0)$, denoted by $\dot{\Psi}(\beta_0,\rho_0,\Lambda_0)$, is continuously invertible.

Then 
\[
\sqrt{m}(\hat{\beta}^T_m-\beta^T_0,\hat{\rho}^T_m-\rho^T_0,\hat{\Lambda}_m-\Lambda_0){\overset{d.}{\rightarrow}}-\dot{\Psi}(\beta_0,\rho_0,\Lambda_0)^{-1}(Z) \ .
\]
\begin{proof}

We shall show that conditions P.1$\sim$P.5 are satisfied.
 
Condition P.1 is shown in Theorem 1.
 
We define a neighbourhood of the true parameter $(\beta_0,\rho_0,\Lambda_0)$, denoted by $U$, a subset of $\Theta$:
\[
U:=\left\{(\beta,\rho,\Lambda):\left|\left|
\beta-\beta_0\right|\right|+
\left|\left|
\rho-\rho_0\right|\right|+
{\underset{t\in [0,\tau]}{\text{sup}}}\left|\Lambda(t)-\Lambda_0(t)\right|<\epsilon_0\right\}\ ,
\]
for a very small fixed constant $\epsilon_0>0$. Clearly, when the sample size $m$ is large enough, $(\hat{\beta}_m,\hat{\rho}_m,\hat{\Lambda}_m)$ belongs to $U$ with probability approaching one. We construct a Banach space to index our infinite-dimensional parameters:
\begin{eqnarray*}
&&\mathcal{H}:=\left\{(h_1,h_2,h_3):  h_1\in\mathbb{R}^{d_1}\text{, }h_2\in\mathbb{R}^{d_2}\text{, }h_3(t) \text{ is a c$\grave{a}$dl$\grave{a}$g function on }[0,\tau]\right\}\ ,\\
&&\text{equipped with the norm }
||h||_{\mathcal{H}}:=||h_1||+||h_2||+||h_3||_{V}\ .
\end{eqnarray*}
Define subspaces of $\mathcal{H}$ $\mathcal{H}_p:=\left\{h\in\mathcal{H}:||h||_{\mathcal{H}}\leq p\right\}$, $\forall p >0$, and the inequality will be strict if $p=\infty$. 

Then we can define $\Psi_m$ and $\Psi$ as maps from $U$ to $l^{\infty}(\mathcal{H}_1)$ such that $l^{\infty}(\mathcal{H}_1)$ consists of all the bounded functions on $\mathcal{H}_{1}$:
\begin{eqnarray*}
\Psi_m(\beta,\rho,\Lambda)[h_1,h_2,h_3]&=& \left.\frac{d}{d t}\mathbb{P}_mclog(\beta+th_1,\rho+th_2,\Lambda+t\int h_3d\Lambda)\right|_{t=0}=\mathbb{P}_mV(\beta,\rho,\Lambda)(h)
\ ,\\
\Psi(\beta,\rho,\Lambda)[h_1,h_2,h_3]&=& \left.\frac{d}{d t}P_0clog(\beta+th_1,\rho+th_2,\Lambda+t\int h_3d\Lambda)\right|_{t=0}=P_0V(\beta,\rho,\Lambda)(h)\ ,\\
\text{where }V(\beta,\rho,\Lambda)[h]&:= & \left.\frac{d}{d t}clog(\beta+th_1,\rho+th_2,\Lambda+t\int h_3d\Lambda)\right|_{t=0}\\ &:= &\left\{
h_1^Tclog_{\beta}(\beta,\rho,\Lambda)
+h_2^Tclog_{\rho}(\beta,\rho,\Lambda)+
clog_{\Lambda}(\beta,\rho,\Lambda)\left[\int h_3d\Lambda\right]
\right\}\ ,
\end{eqnarray*}
and $clog_{\beta}(\beta,\rho,\Lambda)$ is the composite score for $\beta$, $clog_{\rho}(\beta,\rho,\Lambda)$ is the composite score for $\rho$ and $clog_{\Lambda}(\beta,\rho,\Lambda)\left[\int h_3d\Lambda\right]$ is the composite score for $\Lambda$ along the submodel $\Lambda+\epsilon\int h_3d\Lambda$. The equalities are true since the integral and differentiation signs are exchangeable. It is trivial that 
\[
\Psi_m(\hat{\beta}_m,\hat{\rho}_m,\hat{\Lambda}_m)= 0;\quad \Psi(\beta_0,\rho_0,\Lambda_0)= 0\ . 
\]

To show the weak convergence in P.2 of the theorem, we want to verify the function class
\[
\left\{
V(\beta,\rho,\Lambda)(h): (h_1,h_2,h_3)\in\mathcal{H}_1
,(\beta,\rho,\Lambda)\in U\right\}
\]
is $P_0$-Donsker. This procedure is quite similar to the two classes discussed in Appendix B and thus omitted.

To verify P.3, by the $P_0$-Donsker preservation theorem, 
\[
\left\{
V(\beta,\rho,\Lambda)(h)-V(\beta_0,\rho_0,\Lambda_0)(h):(\beta,\rho,\Lambda)\in U, h\in\mathcal{H}_1\right\}
\]
is $P_0$-Donsker as well. Also,
\begin{eqnarray}
&&{\underset{h\in\mathcal{H}}{\text{sup}}}P_0\left[V(\beta,\rho,\Lambda)(h)-V(\beta_0,\rho_0,\Lambda_0)(h)\right]^2\nonumber\\ \leq && P_0\left[M_1 ||\beta_0-\beta||+M_2 ||\rho_0-\rho||+M_3 ||\Lambda_0-\Lambda||_\infty\right]^2\label{survival_normal_jia1}\\
\rightarrow  && 0\quad\text{ as }||(\beta,\rho,\Lambda)-(\beta_0,\rho_0,\Lambda_0)||_\infty\rightarrow 0\ .\nonumber
\end{eqnarray}
\eqref{survival_normal_jia1} is due to the fact that everything in $V(\beta,\rho,\Lambda)(h)$ is continuous with regards to $(\beta,\rho,\Lambda(Y_1),\ldots,\Lambda(Y_n))$ so Mean Value Theorem can be applied. Since all random variables are uniformly bounded, there exists finite constants $(M_1,M_2,M_3)$.

Therefore, according to Lemma 3.3.5 from \citeasnoun{book5}, P.3 holds.

To verify the Fr\'{e}chet differentiability of the composite score function, we first consider the G$\check{a}$teaux derivative of $\Psi$ at $(\beta_0,\rho_0,\Lambda_0)$, denoted by $\dot{\Psi}$, which is a map from the set 
$
\dot{U}\equiv \{(\beta-\beta_0,\rho-\rho_0,\Lambda-\Lambda_0):(\beta,\rho,\Lambda)\in U\}
$ 
to $l^\infty(\mathcal{H}_\infty)$.

Straightforward calculations yield that
\begin{eqnarray*}
&&\dot{\Psi}(\beta-\beta_0,\rho-\rho_0,\Lambda-\Lambda_0)[h_1,h_2,\int h_3d\Lambda_0]\\=&&(\beta-\beta_0)^T\mathcal{T}_{1,\theta_0}(h_1,h_2, h_3)+(\rho-\rho_0)^T\mathcal{T}_{2,\theta_0}(h_1,h_2, h_3)+\int^\tau_0\mathcal{T}_{3,\theta_0}(h_1,h_2, h_3)d(\Lambda-\Lambda_0)
\\=&&(\beta-\beta_0)^T\left[\mathcal{T}_{1,\beta,\theta_0}(h_1)+\mathcal{T}_{1,\rho,\theta_0}(h_2)+\mathcal{T}_{1,\Lambda,\theta_0}( h_3)\right]\\&&+(\rho-\rho_0)^T\left[\mathcal{T}_{2,\beta,\theta_0}(h_1)+\mathcal{T}_{2,\rho,\theta_0}(h_2)+\mathcal{T}_{2,\Lambda,\theta_0}(h_3)\right]\\&& +\int^\tau_0\left[ \mathcal{T}_{3,\beta,\theta_0}(h_1)+\mathcal{T}_{3,\rho,\theta_0}(h_2)+\mathcal{T}_{3,\Lambda,\theta_0}( h_3)\right]d(\Lambda-\Lambda_0)\\=&&\dot{\theta}(\mathcal{T}_{\theta_0}(h))\ .
\end{eqnarray*}
$\dot{\theta}$ is an element from $l^\infty(\mathcal{H}_\infty)$, defined by
\[
\dot{\theta}_1(\mathcal{T}_{\theta_0}(h))=
(\beta_1-\beta_0)^T\mathcal{T}_{1,\theta_0}(h)
+(\rho_1-\rho_0)^T\mathcal{T}_{2,\theta_0}(h)+\int_0^\tau \mathcal{T}_{3,\theta_0}(h)d(\Lambda_1-\Lambda_0)\ ,
\]
where the operator $\mathcal{T}_{\theta_0}:\mathcal{H}_\infty\mapsto\mathcal{H}_\infty$ can be written as
\begin{equation}\label{survival_normal_frechet}
\mathcal{T}_{\theta_0}(h)=
\left(\ba{ccc}\mathcal{T}_{1,\beta,\theta_0}&\mathcal{T}_{1,\rho,\theta_0}&\mathcal{T}_{1,\Lambda,\theta_0}\\ 
\mathcal{T}_{2,\beta,\theta_0}&\mathcal{T}_{2,\rho,\theta_0}&\mathcal{T}_{2,\Lambda,\theta_0}\\ 
\mathcal{T}_{3,\beta,\theta_0}&\mathcal{T}_{3,\rho,\theta_0}&\mathcal{T}_{3,\Lambda,\theta_0}\ea\right)
\left(\ba{c}h_1\\ h_2\\ h_3\ea\right)\ .
\end{equation}
Since integral under $P_0$ and differentiation are exchangeable, we get
\begin{eqnarray*}
\mathcal{T}_{1,\beta,\theta_0}(h_1)&=&h_1^TP_0clog_{\beta\beta}(X;\beta_0,\rho_0,\Lambda_0)\\
\mathcal{T}_{1,\rho,\theta_0}(h_2)&=&h_2^TP_0clog_{\beta\rho}(X;\beta_0,\rho_0,\Lambda_0)\\
\mathcal{T}_{1,\Lambda,\theta_0}(h_3)&=&P_0\int^\tau_0 C_{\beta}(u,X;\beta_0,\rho_0,\Lambda_0)h_3(u)d\Lambda_0(u)\\
\mathcal{T}_{2,\beta,\theta_0}(h_1)&=&h_1^TP_0clog_{\rho\beta}(X;\beta_0,\rho_0,\Lambda_0)\\
\mathcal{T}_{2,\rho,\theta_0}(h_2)&=&h_2^TP_0clog_{\rho\rho}(X;\beta_0,\rho_0,\Lambda_0)\\
\mathcal{T}_{2,\Lambda,\theta_0}(h_3)&=&P_0\int^\tau_0 C_{\rho}(u,X;\beta_0,\rho_0,\Lambda_0)h_3(u)d\Lambda_0(u)\\
\mathcal{T}_{3,\beta,\theta_0}(h_1)(t)&=&P_0\left[\underset{j=1}{\overset{n}{\sum}}q_j(X;\beta_0,\rho_0,\Lambda_0)I(Y_j\geq t)\right]^Th_1\\
\mathcal{T}_{3,\rho,\theta_0}(h_2)(t)&=&P_0\left[\underset{j=1}{\overset{n}{\sum}}\tilde{q}_j(X;\beta_0,\rho_0,\Lambda_0)I(Y_j\geq t)\right]^Th_2\\ 
\mathcal{T}_{3,\Lambda}(h_3)(t)&=&P_0\left(
\int^\tau_0\overset{n}{\underset{ j=1}{\sum}}B_j(u,X;\beta_0,\rho_0,\Lambda_0)1\{Y_j\geq t\}h_3(u)d\Lambda_0(u)-W(t,X;\beta_0,\rho_0,\Lambda_0)h_3(t)\right)\ ,
\end{eqnarray*}

\begin{eqnarray*}
\text{where we define}&&\\
C_{\beta}(u,X;\beta_0,\rho_0,\Lambda_0)&:=&-\bigtriangledown _{\beta}
W(u,X;\beta_0,\rho_0,\Lambda_0)\\
C_{\rho}(u,X;\beta_0,\rho_0,\Lambda_0)&:=&-\bigtriangledown _{\rho}
W(u,X;\beta_0,\rho_0,\Lambda_0)\\
q_j(X;\beta_0,\rho_0,\Lambda_0)&:=&\frac{\partial clog_{\beta}(X;\beta_0,\rho_0,\Lambda_0)}{\partial\Lambda_0(Y_j)}\\
\tilde{q}_j(X;\beta_0,\rho_0,\Lambda_0)&:=&\frac{\partial clog_{\rho}(X;\beta_0,\rho_0,\Lambda_0)}{\partial\Lambda_0(Y_j)}\\
B_j(u,X;\beta_0,\rho_0,\Lambda_0)&:=&-\partial\left(\frac{1}{n-1}{\underset{ k\neq j}{\sum}}\left[(1+\Delta_{j}+\Delta_{k})
\frac{e^{Z_{j}^T\beta_0}1\{Y_j\geq u\}\left[1+(1-\rho_{jk})e^{Z_k^T\beta_0}\Lambda_0(Y_k)\right]}{v(X_{j},X_{k};\beta_0,\Lambda_0,\rho_0)}
\right.\right.\\&&\left.\left.\left.-\frac{\Delta_{k}(1-\rho_{jk}) e^{Z_{j}^T\beta_0}1\{Y_{j}\geq u\}
\left[
\Delta_{j}(1-\rho_{jk})e^{Z_{k}^T\beta_0}\Lambda_0(Y_{k})+1\right]}{w(X_{j},X_{k};\beta_0,\Lambda_0,\rho_0)}\right]\right)\right/\partial \Lambda_0(Y_j)\ .
\end{eqnarray*}
Thus, Fr\'{e}chet differentiability is shown by definition.

In the following we first show $\mathcal{T}_{\theta_0}(h)$ is invertible and then we show it is also a Fredholm operator.

Note that 
\begin{eqnarray*}
&&\mathcal{T}_{\theta_0}(h)\\ \propto &&-\mathrm{E}\left[\frac{1}{n-1}\underset{j=1}{\overset{n}{\sum}}{\underset{j<k}{\sum}}\left\{h_1^Tclog_{\beta}(X_j,X_k;\beta_0,\rho_0,\Lambda_0)+
h_2^Tclog_{\rho}(X_j,X_k;\beta_0,\rho_0,\Lambda_0)\right.\right.\\&&\left.\left.\quad\quad\quad +clog_\Lambda(X_j,X_k;\beta_0,\rho_0,\Lambda_0)[\int h_3d\Lambda_0]\right\}^2\right]\\=&&0\ ;
\end{eqnarray*}
i.e. all pairwise score functions will be zero a.s. for the one-dimensional sub-model defined in the direction of $h$. By Lemma 1, it implies $h=0$ and thus $\mathcal{T}_{\theta_0}(h)$ is invertible.

To show it is a Fredholm operator, we define 
\begin{eqnarray*}
A(h)&:=&\left(\ba{ccc}P_0clog_{\beta,\beta}(\beta_0,\rho,\Lambda_0)&P_0clog_{\beta,\rho}(\beta_0,\rho,\Lambda_0)&0\\P_0clog_{\rho,\beta}(\beta_0,\rho,\Lambda_0)&P_0clog_{\beta,\beta}(\beta_0,\rho,\Lambda_0)&0\\0&0&-P_0W(t;\beta_0,\rho,\Lambda_0)\ea\right)\left(\ba{c}h_1\\ h_2\\ h_3\ea\right)\\&:=&\left(\ba{cc}A_0&0\\0&-P_0W(t;\beta_0,\rho,\Lambda_0)\ea\right)\left(\ba{c}(h_1, h_2)^T\\ h_3\ea\right)\ .
\end{eqnarray*}
$A(h)$ is a continuously invertible operator trivially:
\[
A^{-1}(h)=\left(\ba{cc}A_0^{-1}&0\\0&-\frac{1}{P_0W(t;\beta_0,\rho,\Lambda_0)}\ea\right)\left(\ba{c}(h_1, h_2)^T\\ h_3\ea\right)
\]
We need to show the remaining part $K(h):=\mathcal{T}_{\theta_0}(h)-A(h)$ is a compact operator. We write out $K(h)$ explicitly
\begin{eqnarray*}
K(h)&=& K_1(h)+K_2(h)+K_3(h)\ ,\\ \text{where}&&\\
K_1(h)&=&P_0\int^\tau_0 C_{\beta}(u,X;\beta_0,\rho_0,\Lambda_0)h_3(u)d\Lambda_0(u)+
P_0\int^\tau_0 C_{\rho}(u,X;\beta_0,\rho_0,\Lambda_0)h_3(u)d\Lambda_0(u)\ ,\\
K_2(h)&=&
P_0\left[\underset{j=1}{\overset{n}{\sum}}q_j(X;\beta_0,\rho_0,\Lambda_0)I(Y_j\geq t)\right]^Th_1+
P_0\left[\underset{j=1}{\overset{n}{\sum}}\tilde{q}_j(X;\beta_0,\rho_0,\Lambda_0)I(Y_j\geq t)\right]^Th_2\ ,\\
K_3(h)&=&P_0\left(
\int^\tau_0\overset{n}{\underset{ j=1}{\sum}}B_j(u,X;\beta_0,\rho_0,\Lambda_0)1\{Y_j\geq t\}h_3(u)d\Lambda_0(u)\right)\ .
\end{eqnarray*}
We can see $K_1(h)$ and $K_2(h)$ are  bounded linear operators with finite-dimensional range and thus are compact as in \citeasnoun{ref90}.

For $K_3(h)$, we consider a sequence of indexing elements $\{h_{1n},h_{2n},h_{3n}\}$ such that $||h_{1n}||+||h_{2n}||+||h_{3n}||_{V}\leq 1$. We write every $h_{3n}$ in the form
\begin{eqnarray*}
h_{3n}(t)&=&h_{3n}^+(t)-h_{3n}^-(t)\\
\text{where }h^+_{3n}(t)&=&\left\{\ba{c}h_{3n}(t)\quad\text{if }h_{3n}(t)\geq 0\\ 0\quad\text{otherwise}\ea\right.\\
\text{where }h^-_{3n}(t)&=&\left\{\ba{c}-h_{3n}(t)\quad\text{if }h_{3n}(t)< 0\\ 0\quad\text{otherwise}\ea\right.\ .
\end{eqnarray*}
Since $K_3(h_{3n}^+)(t):=P_0\left(
\int^\tau_0\overset{n}{\underset{ j=1}{\sum}}B_j(u;\beta_0,\rho_0,\Lambda_0)1\{Y_j\geq t\}h_{3n}^+(u)d\Lambda_0(u)\right)$ is a monotone function in $t\in[0,\tau]$ and $B_j(u;\beta_0,\rho_0,\Lambda_0)$ is uniformly bounded, we have
\[
||K_3(h_{3n}^+)||_{V}= |K_3(h_{3n}^+)(0)|\leq C\int^\tau_0h_{3n}^+(u)d\Lambda_0(u)\ .
\]
By Helly's Selection Lemma, there exists a subsequence of $\{h_{3n}\}$:  $\{h_{3n_z}\}$ such that $h_{3n_z}^+\rightarrow g_{03}^+$ and $h_{3n_z}^-\rightarrow g_{03}^-$ point-wise, where $||g_{03}^+||\leq 1$ and $||g_{03}^-||\leq 1$. Then by Dominant Convergence Theorem
\[
||K_3(h_{3n}^+)-K_3(g_{03}^+)||_{V}=||K_3(h_{3n}^+-g_{03}^+)||_{V} \leq C\int^\tau_0\left|h_{3n}^+-g_{03}^+\right|(u)d\Lambda_0(u)\rightarrow 0\ . 
\]
Thus
\[
||K_3(h_{3n})-K_3(g_{03})||_{V}\rightarrow 0 \ .
\]
Therefore we have shown there exists a subsequence and an element $g_0\in\mathcal{H}^\infty$ such that  
\[
||K(h_{n_k})-g_0||\rightarrow 0\ .
\]
As a summary, we have shown the operator $\mathcal{T}_{\theta_0}:\mathcal{H}_\infty\mapsto\mathcal{H}_\infty$ is a Fredholm operator.

Since the operator $\mathcal{T}_{\theta_0}:\mathcal{H}_\infty\mapsto\mathcal{H}_\infty$ is a Fredholm operator and is one-to-one, by Lemma 6.17 in \citeasnoun{book8}, it is continuously invertible and onto.

According to Lemma 6.16 in \citeasnoun{book8}, continuous invertibility of $\mathcal{T}_{\theta_0}(h)$ implies continuous invertibility $\dot{\theta}(\mathcal{T}_{\theta_0}(h))$, if
for each $p>0$, there is a $q>0$ such that $\mathcal{T}^{-1}_{\theta_0}(\mathcal{H}_q)\subset\mathcal{H}_p$, i.e. $\mathcal{T}_{\theta_0}(\mathcal{H}_p)\text{ maps onto }\mathcal{H}_q$. Fix some $p>0$ and use the conclusion from Exercise 15.6.4 in \citeasnoun{book8}:
\[
{\underset{\dot{\theta}\in \text{lin}\dot{U}}{\text{inf}}}\frac{||\dot{\theta}(\mathcal{T}_{\theta_0}(\cdot))||_{(p)}}{||\dot{\theta}(\cdot)||_{(p)}}=
{\underset{\dot{\theta}\in \text{lin}\dot{U}}{\text{inf}}}\frac{{\underset{h\in\mathcal{H}_p}{\text{sup}}}|\dot{\theta}(\mathcal{T}_{\theta_0}(h))|}{{\underset{h\in\mathcal{H}_p}{\text{sup}}}|\dot{\theta}(h)|}
\geq 
{\underset{\dot{\theta}\in \text{lin}\dot{U}}{\text{inf}}}\frac{{\underset{\tilde{h}\in\mathcal{H}_q}{\text{sup}}}|\dot{\theta}(\tilde{h})|}{{\underset{h\in\mathcal{H}_p}{\text{sup}}}|\dot{\theta}(h)|}
\geq\left\{\ba{cc}1&q
\geq p\\ \frac{q}{2p}&q<p\ea\right.\ .\] 
We conclude operator $\dot{\theta}(\mathcal{T}_{\theta_0}(h))$ is continuously invertible.

Therefore, conditions P.1$\sim$P.5 are satisfied and Theorem 2 holds.
\end{proof}
\section*{Appendix B}
Here we show two classes of functions are $P_0$-Donsker and the third class of functions is $P_0$-Glivenko-Cantelli.
\subsection*{\textbf{The first class of functions}}
Remember we defined
\[
G(O;t)=\overset{n}{\underset{j=1}{\sum}}\Delta_j 1\{Y_j\leq t\}\ .
\]
Then following class of functions, indexed by $t\in[0,\tau]$:
\[
\mathcal{F}_2:=\left\{f_2(O;t):=
\int^t_0 g(s)dG(O;s) :g \text{ is a c$\grave{a}$dl$\grave{a}$g function on  $[0,\tau]$ and } ||g||_V\leq M_1<\infty,O\sim P_{0}\right\}
\]
is $P_0$-Donsker.
\begin{proof}
Consider another function class which is also indexed by $t$
\[
\mathcal{F}_0:=\left\{f_0(O;t):=
\int^t_0 g(s)dG(O;s) :g \text{ is monotone on $[0,\tau]$ and } ||g||_V\leq M_1<\infty,O\sim P_{0}\right\}\ ,
\]
which can be rewritten as
\begin{eqnarray*}
\mathcal{F}_0=&&\left\{f_0(O;t):=
\underset{j=1}{\overset{n}{\sum}}g(Y_{j})1\{Y_{j}\leq t\} \Delta_{j}:t\in[0,\tau],\right.\\&&\left.\quad g \text{ is a monotone function on $[0,\tau]$ and } ||g||_V\leq M_1<\infty,O\sim P_{0}\right\}\ .
\end{eqnarray*}
For a single observation denoted by $X_1$, consider the function class:
\begin{eqnarray*}
\mathcal{F}_1=&&\left\{f_1(X_1;t):=
g(Y_1)1\{Y_1\leq t\} \Delta_1:t\in[0,\tau],\right.\\ &&\left.\quad g \text{ is a monotone function on $[0,\tau]$ and } ||g||_V\leq M_1<\infty\right\}\ .
\end{eqnarray*}
For some fixed $X_1$, $f_1(X_1;t)$ is a mono-increasing function in $t$. According to Exercise 3 on page 165 in \citeasnoun{book5}, number of brackets is in a polynomial order and thus $\mathcal{F}_1$ is $P_{0}$-Donsker. That is to say,
for any $\epsilon>0$, denote $[f_1(X_1;t_s^L),f_1(X_1;t_s^U)]$, $s=1,\ldots,N_\epsilon$ as the set of brackets covering $\mathcal{F}_1$ such that
\[
||f_1(X_1;t_s^L)-f_1(X_1;t_s^U)||_{L_2(P_0)}\leq \epsilon/n_0
\]
and $N_\epsilon$ is in the order $(O(1)/\epsilon)^{M_{\epsilon}}$, $M_{\epsilon}$ is some finite number.

For an arbitrary $f_{0}(O;t')$ from $\mathcal{F}_0$, we can find some $s=1,\ldots, N_\epsilon$ such that
\begin{eqnarray*}
&& f_1(X_j;t_s^L)\leq f_1(X_j;t')\leq f_1(X_j;t_s^U),\quad j=1,\ldots,n\\
&&||f_0(O;t_s^L)-f_0(O;t_s^U)||_{L_2(P_0)}\leq  \underset{j=1}{\overset{n}{\sum}}||f_1(X_j;t_s^L)-f_1(X_j;t_s^U)||_{L_2(P_0)}\leq \epsilon\ .
\end{eqnarray*}
Since there is also a square integrable envelop function for $\mathcal{F}_0$, $\mathcal{F}_0$ is $P_0$-Donsker. Since every element in $\mathcal{F}_2$ can be expressed as a summation of two elements in $\mathcal{F}_0$, $\mathcal{F}_2$ is also $P_0$-Donsker by the preservation theorem.
\end{proof}
\subsection*{\textbf{The second class of functions}}

The function class:
\[
\mathcal{W}=\{W(u,O;\beta,\rho,\Lambda):\quad O\sim P_0;\quad u\in[0,\tau],\quad (\beta,\rho,\Lambda)\in\Theta\}
\]
is $P_0$-Donsker, where 
\begin{eqnarray*}
W(u,O;\beta,\rho,\Lambda)&=&\overset{n}{\underset{j=1}{\sum}}
\left\{\frac{1}{n-1}{\underset{k\neq j}{\sum}}\left[(1+\Delta_{j}+\Delta_{k})
\frac{e^{Z_{j}^T\beta}1\{Y_j\geq u\}\left[1+(1-\rho_{jk})e^{Z_k^T\beta}\Lambda(Y_k)\right]}{v(X_{j},X_{k};\beta,\Lambda,\rho)}
\right.\right.\nonumber\\&&\left.\left.-\frac{\Delta_{k}(1-\rho_{jk}) e^{Z_{j}^T\beta}1\{Y_{j}\geq u\}
\left[
\Delta_{j}(1-\rho_{jk})e^{Z_{k}^T\beta}\Lambda(Y_{k})+1\right]}{w(X_{j},X_{k};\beta,\Lambda,\rho)}\right]\right\},\quad O\sim P_0\ .
\end{eqnarray*}
\begin{proof}
For an arbitrary pair of elements from $\mathcal{W}$, given a sample, their absolute difference is bounded by
\[
\left| W(u_1,O;\beta_2,\rho_2,\Lambda_2)-W(u_1,O;\beta_1,\rho_1,\Lambda_1)\right| \leq 
A_0\left\{
||\beta_1-\beta_2|| + ||\rho_1-\rho_2|| +\underset{j=1}{\overset{n}{\sum}} |\Lambda_1(Y_j) -\Lambda_2(Y_j)|
\right\}\ .
\]
The above bound is achieved by noting $W(u,O;\beta,\rho,\Lambda)$ is absolute continuous in $(\beta,\rho,\Lambda(Y_1),\ldots,\Lambda(Y_n))$ and every element in $W(u,O;\beta,\rho,\Lambda)$ is uniformly bounded; thus $W(u,O;\beta,\rho,\Lambda)$ is Lipschitz continuous in $(\beta,\rho,\Lambda(Y_1),\ldots,\Lambda(Y_n))$. Define a function 
\[
h(O;\Lambda_1,\Lambda_2)=\underset{j=1}{\overset{n}{\sum}}| \Lambda_2(Y_j)-\Lambda_1(Y_j)  |\ .
\]
The right side of the above inequality can be rewritten as\\ \noindent $A_0\left\{
||\beta_1-\beta_2|| + ||\rho_1-\rho_2|| +h(O;\Lambda_1,\Lambda_2)
\right\}$.

By Theorem 2.7.5 in \citeasnoun{book5}, we have the number of brackets of 
\begin{eqnarray}
\{&&\Lambda(\cdot):\text{ non-decreasing step function in }[0,\tau]\text{ with jumps at the observed failure times}\nonumber\\&&\text{ and }\Lambda(0)=0,\Lambda(\tau)\leq C\} \nonumber
\end{eqnarray}
in the order of $exp(O(1)/\epsilon)$, under probability measure $P_0$. Due to the compactness in finite-dimensional part of the parameter, number of brackets on $\Theta$ is in the order $exp(O(1)/\epsilon)$. By definition, we have for $\forall\epsilon >0$, there exists a finite bracket interval $[\beta_s^L,\beta_s^U]\times [\rho_s^L,\rho_s^U]\times [\Lambda_s^L,\Lambda_s^U]$, $s=1,\ldots,N_\epsilon$ covering $\Theta$, where $N_\epsilon\sim exp(O(1)/\epsilon)$ such that for arbitrary $\theta'=(\beta',\rho',\Lambda')\in\Theta$, there is some $s$ such that 
\begin{eqnarray*}
&&\beta_s^L\leq \beta'\leq\beta_s^U ,\quad \rho_s^L \leq\rho' \leq\rho_s^U,\quad \Lambda_s^L(\cdot)\leq \Lambda'(\cdot) \leq\Lambda_s^U(\cdot);
\\&&||\beta_s^L-\beta_s^U||<\frac{\epsilon}{12n_0A_0},\quad 
||\rho_s^L-\rho_s^U||<\frac{\epsilon}{12n_0A_0},\quad 
||\Lambda_s^L(Y)-\Lambda_s^U(Y)||_{L_2(P_0)}<\frac{\epsilon}{12n_0A_0}\ ;\\
\text{i.e.}&& ||\beta_s^L-\beta'||\leq ||\beta_s^L-\beta_s^U||<\frac{\epsilon}{12n_0A_0},\quad 
||\rho_s^L-\rho'||\leq ||\rho_s^L-\rho_s^U||<\frac{\epsilon}{12n_0A_0},
\\&&h(O;\Lambda_s^L,\Lambda') \leq  h(O;\Lambda_s^L,\Lambda_s^U),\\&&||h(O;\Lambda_s^L,\Lambda')||_{L_2(P_0)}\leq ||h(O;\Lambda_s^L,\Lambda_s^U)||_{L_2(P_0)}<\frac{\epsilon}{12A_0}\ .
\end{eqnarray*}
Fixing bracketing set and consider the function classes, $s=1,\ldots,N_\epsilon$:
\[
\mathcal{W}_{\epsilon,s}:=\left\{ W(u,O;\beta_s^L,\rho_s^L,\Lambda_s^L): u\in[0,\tau] \right\}\ .
\]
By Exercise 3 on page 165 in \citeasnoun{book5}, we have the number of brackets for $\mathcal{W}_{\epsilon,s}$, denote by
$[u^L_{\epsilon,s,t},u^U_{\epsilon,s,t}]$, $t=1,\ldots N_\epsilon^{\epsilon,s}$, where $N_\epsilon^{\epsilon,s}$ is in a polynomial order. That is to say, for arbitrary $\forall u\in[0,\tau]$, there is some $t=1,\ldots,N_\epsilon^{\epsilon,s}$ such that
\begin{eqnarray*}
&&W(u^L_{\epsilon,s,t},O;\beta_s^L,\rho_s^L,\Lambda_s^L)\leq W(u,O;\beta_s^L,\rho_s^L,\Lambda_s^L)\leq W(u^U_{\epsilon,s,t},O;\beta_s^L,\rho_s^L,\Lambda_s^L)\ ,
\\&&||W(u^L_{\epsilon,s,t},O;\beta_s^L,\rho_s^L,\Lambda_s^L)-W(u^U_{\epsilon,s,t},O;\beta_s^L,\rho_s^L,\Lambda_s^L)||_{L_2(P_0)}<\frac{\epsilon}{2}\ .
\end{eqnarray*}
For an arbitrary function $W(u',O;\beta',\rho',\Lambda')\in\mathcal{W}$, it is contained in bracket
\begin{eqnarray*}
&& \left[ W(u^L_{\epsilon,s,t},O;\beta_s^L,\rho_s^L,\Lambda_s^L)-A_0\left\{ ||\beta_s^U-\beta_s^L|| +||\rho_s^U-\rho_s^L|| + 
h(O;\Lambda_s^L,\Lambda_s^U)\right\},\right. \\ \quad\quad &&
\left. \quad W(u^U_{\epsilon,s,t},O;\beta_s^L,\rho_s^L,\Lambda_s^L)+A_0\left\{ ||\beta_s^U-\beta_s^L|| + ||\rho_s^U-\rho_s^L|| + 
h(O;\Lambda_s^L,\Lambda_s^U)\right\}\right]\ ,
\end{eqnarray*}
such that the distance between the boundary functions:
\begin{scriptsize}
\begin{eqnarray*}
&&||W(u^U_{\epsilon,s,t},O;\beta_s^L,\rho_s^L,\Lambda_s^L) -W(u^L_{\epsilon,s,t},O;\beta_s^L,\rho_s^L,\Lambda_s^L)+2A_0\left\{ ||\beta_s^U-\beta_s^L|| +||\rho_s^U-\rho_s^L|| + 
h(O;\Lambda_s^L,\Lambda_s^U)\right\}||_{L_2(P_0)}\\
= && ||W(u^U_{\epsilon,s,t},O;\beta_s^L,\rho_s^L,\Lambda_s^L) -W(u^L_{\epsilon,s,t},O;\beta_s^L,\rho_s^L,\Lambda_s^L)+2A_0\left\{ ||\beta_s^U-\beta_s^L|| +||\rho_s^U-\rho_s^L|| + 
h(O;\Lambda_s^L,\Lambda_s^U)\right\}||_{L_2(P_0)}\\
\leq && ||W(u^L_{\epsilon,s,t},O;\beta_s^L,\rho_s^L,\Lambda_s^L)-W(u^U_{\epsilon,s,t},O;\beta_s^L,\rho_s^L,\Lambda_s^L)||_{L_2(P_0)}\\&&+
2A_0||\beta_s^L-\beta_s^U||+2A_0||\rho_s^L-\rho_s^U|| +2A_0  ||h(O;\Lambda_s^L,\Lambda_s^U)||_{L_2(P_0)}
\\
\leq && \frac{\epsilon}{2} +\frac{\epsilon}{6} +\frac{\epsilon}{6} +\frac{\epsilon}{6}<\epsilon \ .
\end{eqnarray*}
\end{scriptsize}
Since it is straightforward to show the number of brackets is in the order $exp(O(1)/\epsilon)$, thus by definition, $\mathcal{W}$ is $P_0$-Donsker.
\end{proof}

\subsection*{\textbf{The third class of functions}}
The function sequence indexed by the estimator sequence $\left\{\left(\hat{\beta}_{n'},\hat{\rho}_{n'},\hat{\Lambda}_{n'}\right),\left(\beta_0,\rho_0,\tilde{\Lambda}_{n'}\right) \right\}$:
\[
\{clog(O;\hat{\beta}_{n'},\hat{\rho}_{n'},\hat{\Lambda}_{n'})-clog(O;\beta_0,\rho_0,\tilde{\Lambda}_{n'});\, O\sim P_0\}
\]
is $P_0$-Glivenko-Cantelli.
\begin{proof}
First, we want to show this sequence is contained in 
function class $\mathcal{G}_0$ defined by:
\begin{eqnarray*}
\mathcal{G}_0&=&\{f(O;\beta_1,\rho_1,\Lambda_1,\beta_2,\rho_2,\Lambda_2)=clog(O;\beta_1,\rho_1,\Lambda_1)-clog(O;\beta_2,\rho_2,\Lambda_2):\\&&\quad (\beta_1,\rho_1,\Lambda_1),(\beta_2,\rho_2,\Lambda_2)\in\Theta,y\mapsto\frac{\Delta \Lambda_1}{\Delta \Lambda_2}(u)\in [m_1,M_1]\text{ and is }BV_{M_2};O\sim P_0\}\ .
\end{eqnarray*}
For an arbitrary dataset containing $m$ independent clusters, we have this relationship
\[
\frac{\Delta\hat{\Lambda}_m}{\Delta\tilde{\Lambda}_m}(u)=\frac{W_m(u;\beta_0,\rho_0,\Lambda_0)}{W_m(u;\hat{\beta}_m,\hat{\rho}_m,\hat{\Lambda}_m)}\ .
\]
Consider the partition at observed failure event time points $t_1<t_2<\ldots<t_Q$, and set $t_0=0$, $t_{Q+1}=\tau$, then we can write the total variation of $\frac{\Delta\hat{\Lambda}_m}{\Delta\tilde{\Lambda}_m}$ as:
\begin{small}
\begin{eqnarray}
&&\underset{q=0}{\overset{Q}{\sum}}\left|\frac{W_m(t_{q+1};\beta_0,\rho_0,\Lambda_0)}{W_m(t_{q+1};\hat{\beta}_m,\hat{\rho}_m,\hat{\Lambda}_m)}-\frac{W_m(t_{q};\beta_0,\rho_0,\Lambda_0)}{W_m(t_{q};\hat{\beta}_m,\hat{\rho}_m,\hat{\Lambda}_m)}\right| \nonumber\\=&&
\underset{q=0}{\overset{Q}{\sum}}\frac{\left|
W_m(t_{q+1};\beta_0,\rho_0,\Lambda_0)W_m(t_{q};\hat{\beta}_m,\hat{\rho}_m,\hat{\Lambda}_m)-W_m(t_{q};\beta_0,\rho_0,\Lambda_0)W_m(t_{q+1};\hat{\beta}_m,\hat{\rho}_m,\hat{\Lambda}_m)
\right|}{W_m(t_{q+1};\hat{\beta}_m,\hat{\rho}_m,\hat{\Lambda}_m)\cdot W_m(t_{q};\hat{\beta}_m,\hat{\rho}_m,\hat{\Lambda}_m)} \nonumber\\ \leq &&
\underset{q=0}{\overset{Q}{\sum}}\frac{\left|
W_m(t_{q+1};\beta_0,\rho_0,\Lambda_0)W_m(t_{q};\hat{\beta}_m,\hat{\rho}_m,\hat{\Lambda}_m)-W_m(t_{q};\beta_0,\rho_0,\Lambda_0)W_m(t_{q+1};\hat{\beta}_m,\hat{\rho}_m,\hat{\Lambda}_m)
\right|}{m_2^2} \label{survival_consistent_tmp4}
\end{eqnarray}
\end{small}
where $m_2$ denotes the lower bound of $W(\cdot;\cdot,\cdot,\cdot)$.

For some $0\leq q < Q$, suppose $\Delta_{ij}Y_{ij}\in(t_{q},t_{q+1}]$, then
\begin{eqnarray*}
&&W_m(t_{q};\beta,\rho,\Lambda)-W_m(t_{q+1};\beta,\rho,\Lambda)\\=&&\frac{1}{m}\overset{m}{\underset{i=1}{\sum}}\frac{1}{n_i-1}\overset{n_i}{\underset{j=1}{\sum}}{\underset{k\neq j}{\sum}}
\left\{(2+\Delta_{ik})
\frac{e^{Z_{ij}^T\beta}\left[1+(1-\rho_{jk})e^{Z_{ik}^T\beta}\Lambda(Y_k)\right]}{v(X_{ij},X_{ik};\beta,\Lambda,\rho)}
\right.\nonumber\\&&\left.-\frac{\Delta_{ik}(1-\rho_{jk}) e^{Z_{ij}^T\beta}
\left[
(1-\rho_{jk})e^{Z_{ik}^T\beta}\Lambda(Y_{ik})+1\right]}{w(X_{ij},X_{ik};\beta,\Lambda,\rho)}\right\}\\ \leq &&
\frac{1}{m_2}M_2
\end{eqnarray*}
uniformly for $(\beta,\rho,\Lambda)$ varying over $\Theta$.

Thus \eqref{survival_consistent_tmp4} is bounded from above by some constant not changing with $m$, and thus 
each element of estimator sequence: $\left\{\left(\hat{\Lambda}_m,\tilde{\Lambda}_m\right)\right\}$ belong to the set of pairwise estimators:
\[
\left\{(\Lambda_1,\Lambda_2): \Lambda_1\in\mathcal{L},\Lambda_2\in\mathcal{L},y\mapsto\frac{\Delta \Lambda_1}{\Delta \Lambda_2}(u)\in [m_1,M_1]\text{ and is }BV_{M_2}\right\}
\]
Then we want to show $\mathcal{G}_0$ is $P_0$-Glivenko-Cantelli. We write functions from $\mathcal{G}_0$ as
\begin{scriptsize}
\begin{eqnarray}
&&clog(O;\beta_1,\rho_1,\Lambda_1)-clog(O;\beta_2,\rho_2,\Lambda_2)\nonumber\\=&&\overset{n}{\underset{j=1}{\sum}}\Delta_j\left\{ log\frac{\Delta\Lambda_1(Y_j)}{\Delta\Lambda_2(Y_j)}\right\}+
\overset{n}{\underset{j=1}{\sum}}Z_j^T(\beta_1-\beta_2)
\nonumber\\&&\quad -\frac{1}{n-1}{\underset{j<k}{\sum}}\left[(1+\Delta_j+\Delta_k)log\frac{(1-\rho_{1jk})e^{Z_j^T\beta_1}e^{Z_k^T\beta_1}
\Lambda_1(Y_j)\Lambda_1(Y_k)+e^{Z_j^T\beta_1}\Lambda_1(Y_j)+
e^{Z_k^T\beta_1}\Lambda_1(Y_k)+1}{(1-\rho_{2jk})e^{Z_j^T\beta_2}e^{Z_k^T\beta_2}
\Lambda_2(Y_j)\Lambda_2(Y_k)+e^{Z_j^T\beta_2}\Lambda_2(Y_j)+
e^{Z_k^T\beta_2}\Lambda_2(Y_k)+1}\right]
\nonumber\\&&\quad +\frac{1}{n-1}{\underset{j<k}{\sum}}\nonumber\\&&\quad \quad \frac{\Delta_j\Delta_k(1-\rho_{1jk})^2e^{Z_j^T\beta_1}e^{Z_k^T\beta_1}
\Lambda_1(Y_j)\Lambda_1(Y_k)+\Delta_k(1-\rho_{1jk})e^{Z_j^T\beta_1}\Lambda_1(Y_j)
+\Delta_j(1-\rho_{1jk})e^{Z_k^T\beta_1}\Lambda_1(Y_k)+1+\Delta_j\Delta_k\rho_{1jk}
}{\Delta_j\Delta_k(1-\rho_{2jk})^2e^{Z_j^T\beta_2}e^{Z_k^T\beta_2}
\Lambda_2(Y_j)\Lambda_2(Y_k)+\Delta_k(1-\rho_{2jk})e^{Z_j^T\beta_2}\Lambda_2(Y_j)
+\Delta_j(1-\rho_{2jk})e^{Z_k^T\beta_2}\Lambda_2(Y_k)+1+\Delta_j\Delta_k\rho_{2jk}
}\nonumber\ . \\&& \label{survival_consistent_tmp0}
\end{eqnarray}
\end{scriptsize}
The first term in the above is $P_0$-Glivenko-Cantelli with a similar argument as for $\mathcal{F}_2$. The remaining terms form a Lipschitz function of $(\beta_1,\rho_1,\Lambda_1(Y_1),\ldots,\Lambda_1(Y_n),\beta_2,\rho_2,\Lambda_2(Y_1),\ldots,\Lambda_2(Y_n))$ and similar to $\mathcal{W}$ argument, they also form a $P_0$-Glivenko-Cantelli class of functions. By addition preservation Corollary 9.27 in \citeasnoun{book8}, we have shown $\mathcal{G}_0$ is $P_0$-Glivenko-Cantelli.
\end{proof}

\label{lastpage}

\end{document}